\documentclass[10pt]{article}

\usepackage{fullpage}
\usepackage{setspace}
\usepackage{parskip}
\usepackage{titlesec}
\usepackage[section]{placeins}
\usepackage{xcolor}
\usepackage{breakcites}
\usepackage{lineno}
\usepackage{hyphenat}

\PassOptionsToPackage{hyphens}{url}
\usepackage[colorlinks = true,
            linkcolor = blue,
            urlcolor  = blue,
            citecolor = blue,
            anchorcolor = blue]{hyperref}
\usepackage{etoolbox}

\renewenvironment{abstract}
  {{\bfseries\noindent{\abstractname}\par\nobreak}\footnotesize}
  {\bigskip}

\titlespacing{\section}{0pt}{*3}{*1}
\titlespacing{\subsection}{0pt}{*2}{*0.5}
\titlespacing{\subsubsection}{0pt}{*1.5}{0pt}

\usepackage{authblk}

\usepackage{graphicx}
\usepackage[space]{grffile}
\usepackage{latexsym}
\usepackage{textcomp}
\usepackage{longtable}
\usepackage{tabulary}
\usepackage{booktabs,array,multirow}
\usepackage{amsfonts,amsmath,amssymb}
\providecommand\citet{\cite}
\providecommand\citep{\cite}

\newif\iflatexml\latexmlfalse

\AtBeginDocument{\DeclareGraphicsExtensions{.pdf,.PDF,.eps,.EPS,.png,.PNG,.tif,.TIF,.jpg,.JPG,.jpeg,.JPEG}}

\usepackage[utf8]{inputenc}
\usepackage[english]{babel}

\usepackage{float}

\usepackage[margin=1.5in]{geometry}

\usepackage[english]{babel}
\usepackage{listings}
\usepackage{xcolor}
\usepackage{subcaption} % For subfigures

\definecolor{codegreen}{rgb}{0,0.6,0}
\definecolor{codegray}{rgb}{0.5,0.5,0.5}
\definecolor{codepurple}{rgb}{0.58,0,0.82}
\definecolor{backcolour}{rgb}{0.95,0.95,0.92}

\lstdefinestyle{mystyle}{
  backgroundcolor=\color{backcolour},   commentstyle=\color{codegreen},
  keywordstyle=\color{magenta},
  numberstyle=\tiny\color{codegray},
  stringstyle=\color{codepurple},
  basicstyle=\ttfamily\footnotesize,
  breakatwhitespace=false,
  breaklines=true,
  captionpos=b,
  keepspaces=true,
  numbers=left,
  numbersep=5pt,
  showspaces=false,
  showstringspaces=false,
  showtabs=false,
  tabsize=2
}

\usepackage{amsmath}
\usepackage{graphicx}
\usepackage{multirow}
\usepackage{authblk}
\usepackage[utf8]{inputenc}

\usepackage{authblk}
\makeatletter
\renewcommand\AB@affilsepx{ \protect\Affilfont}
\makeatother

\begin{document}
%\title{Advances of Quantum Chemistry Calculation with GPU Acceleration}
\title{Enhancing GPU-acceleration in the Python-based Simulations of Chemistry Framework}
%\title{}
%\title{Python-Based Quantum Chemistry Calculations with GPU Acceleration}
\author[1,*]{Xiaojie Wu}
\author[1,**]{Qiming Sun}
\author[3]{Zhichen Pu}
\author[3]{Tianze Zheng}
\author[3]{Wenzhi Ma}
\author[1]{Wen Yan}
\author[3]{Yu Xia}
\author[3,$\ddagger$]{Zhengxiao Wu}
\author[3,$\ddagger$]{Mian Huo}
\author[3]{Xiang Li}
\author[3]{Weiluo Ren}
\author[1]{Sheng Gong}
\author[1]{Yumin Zhang}
\author[3]{Weihao Gao}

\affil[1]{ByteDance Research, USA\newline}
\affil[3]{ByteDance Research, China\newline\newline}

\affil[*]{xiaojie.wu@bytedance.com\newline}
\affil[**]{osirpt.sun@gmail.com\newline}
\affil[$\ddagger$]{work done at ByteDance as intern}

% ----------------- WIREs begin -------------------------
\vspace{-1em}
\date{}
\begingroup
\let\center\flushleft
\let\endcenter\endflushleft
\maketitle
\endgroup
% --------------------- WIREs end --------------------------
%\footnotetext{work done at ByteDance}

%\maketitle    % uncomment this for arxiv
\selectlanguage{english}
\begin{abstract}
    We describe our contribution as industrial stakeholders to the existing open-source GPU4PySCF project (\url{https://github.com/pyscf/gpu4pyscf}), a GPU-accelerated Python quantum chemistry package. We have integrated GPU acceleration into other PySCF functionality including Density Functional Theory (DFT), geometry optimization, frequency analysis, solvent models, and density fitting technique.
    Through these contributions, GPU4PySCF v1.0 can now be regarded as a fully functional and industrially relevant platform which we demonstrate in this work through a range
of tests.
    When performing DFT calculations on modern GPU platforms,
    GPU4PySCF delivers 30 times speedup over a 32-core CPU node, resulting
    in approximately 90\% cost savings for most DFT tasks. The performance advantages
    and productivity improvements have been found in multiple industrial
    applications, such as generating potential energy surfaces, analyzing molecular
    properties, calculating solvation free energy, identifying chemical reactions in
    lithium-ion batteries, and accelerating neural-network methods. 
    With the improved design that makes it easy to integrate with the Python and PySCF ecosystem, GPU4PySCF is natural choice that we can now recommend for many industrial quantum chemistry applications.
\end{abstract}

\section{Introduction}
Quantum chemistry is essential in various fields such as drug discovery, materials science, chemical engineering, and environmental science. It primarily relies on {\it ab initio} simulations for atomic interactions, serving as a crucial source of data beyond experimental findings. These fields often require extensive ab initio simulations to explore chemical spaces, identify equilibrium geometries, and determine chemical properties. Density Functional Theory (DFT) stands out as the most commonly used method in quantum chemistry for generating physical observables. According to a 2018 workload analysis by the National Energy Research Scientific Computing Center (NERSC), DFT algorithms account for 30\% of total computational resource usage\footnote{\url{https://portal.nersc.gov/project/m888/nersc10/workload/N10_Workload_Analysis.latest.pdf}}. The surge in demand for quantum chemistry calculations is not only due to traditional scientific simulations but also the emergence of deep learning models. These models significantly increase the demand for data from ab initio calculations and use quantum features as input. Consequently, the development of DFT algorithms for GPUs has become increasingly beneficial for these industries.

%Quantum chemistry plays a crucial role in various fields, including drug discovery, materials science, chemical engineering, and environmental science. The ab initio simulation of atomic interactions is the main source of reliable data other than experiments. Vast amounts of ab initio simulations are often required in those fields to sample the chemical space, find the equilibrium geometry, and calculate chemical properties. Density functional theory (DFT) is the most widely used quantum chemistry method to produce physical observables. The algorithms of DFT have been extensively explored in the high-performance computing community. According to NERSC's workload analysis in 2018, DFT algorithms takes 30\% of the total computational resources \footnote{\url{https://portal.nersc.gov/project/m888/nersc10/workload/N10_Workload_Analysis.latest.pdf/}}.  Besides the traditional scientific simulations, the recent deep learning models also bump up the demands of quantum chemistry calculations. On the one hand those models consume a huge amount of data by ab initio calculations. On the other hand, they take the quantum features as the input. The development of the DFT algorithms on GPU practically benefits those industries.

The utilization of GPUs in quantum chemistry has emerged as a transformative tool, thanks to their parallel processing capabilities. GPUs are adept at managing large data volumes and conducting calculations across multiple cores simultaneously, making them exceptionally efficient for quantum chemistry simulations. The initiative to accelerate two-electron integral evaluations using GPUs can be traced back to 2008~\cite{qc-gpu0, qc-gpu1}, a period when gaming cards exhibited limited double-precision computing capabilities. Today's GPUs not only offer substantial floating-point performance but also boast high memory bandwidth and extensive memory capacity, effectively minimizing the bottlenecks associated with quantum chemistry algorithms. Traditional CPU-based quantum chemistry software packages, such as GAMESS~\cite{GAMESS, GAMESS_GPU}, GAUSSIAN, Q-Chem~\cite{qchem}, and Psi4~\cite{psi4} now delegate computationally intensive tasks to GPUs through interfaces like BrainQC~\cite{brainqc1,brainqc2}. Furthermore, the past decade has seen the development of GPU-based quantum chemistry packages, including Terachem~\cite{terachem, terachem_cloud} and Quick~\cite{quick1, quick2, quick3}, which notably reduce CPU-GPU data transfers.

%Graphics Processing Unit (GPU) computing has emerged as a powerful tool in quantum chemistry due to its parallel capabilities. GPUs are designed to handle large amounts of data and perform calculations simultaneously across multiple cores, which make them highly efficient for quantum chemistry simulations. The idea of accelerating two-electron integral evaluation with GPUs can be traced back to 2008~\cite{qc-gpu0,qc-gpu1}, when the gaming cards with limited double-precision computing capability. Modern GPUs are not only delivering high levels of floating-point performance, but also have high memory bandwidth and large memory capacity. The high-performed memory is minimizing the bottleneck of quantum chemistry algorithms. The traditional CPU-based quantum chemistry packages also offload the expensive tasks to GPU, such as GAMESS~\cite{GAMESS,GAMESS_GPU}, GAUSSIAN, Q-Chem~\cite{qchem} and Psi4~\cite{psi4} via BrainQC~\cite{brainqc1,brainqc2}. In the last decade, pure GPU-based quantum chemistry packages such as Terachem~\cite{terachem, terachem_cloud}, Quick~\cite{quick1, quick2, quick3} are also developed where the CPU-GPU data transfer is significantly eliminated.

Achieving optimal performance in quantum chemistry computations on Graphics Processing Units (GPUs) necessitates an approach that extends significantly beyond the simple porting of algorithms from Central Processing Units (CPUs) to GPU architectures. This complex process is hindered by several technical challenges that can substantially impact the acceleration of quantum chemistry simulations on GPUs. 1) {\it Limited memory}. GPUs are equipped with high-speed memory that, despite its rapid access capabilities, is limited in capacity and substantially smaller than the system RAM found in CPU-based systems. The quantum chemistry domain, characterized by its reliance on large matrices and complex wave functions, frequently demands more memory than GPUs can provide. This limitation often requires computational strategies such as offloading portions of the data back to the CPU or the implementation of advanced memory management techniques, both of which can negatively affect computational throughput. 2) {\it CPU-GPU communication}. The bandwidth between CPU memory and GPU memory is restricted and can become a critical bottleneck, especially in computational scenarios where the GPU necessitates frequent access or updates to data stored in the main system memory. This bottleneck necessitates the re-implementation of a significant number of traditional CPU algorithms directly on the GPU to minimize data transfer overhead. 3) {\it Complexities of GPU architecture}. The process of maximizing GPU computational efficiency involves the fine-tuning of several coding and architectural aspects, including but not limited to, optimizing memory access patterns, managing thread execution, and configuring kernels. The same optimization strategies do not necessarily apply to the new architecture. This level of optimization is highly specialized and demands a considerable investment of time, highlighting the intricate and expert-driven nature of leveraging GPU capabilities for the advancement of quantum chemistry calculations.

%An efficient solution of quantum chemistry calculations on GPU is more than reimplementing the classical CPU algorithms on GPU devices. Various technical difficulties can slow down the accelerated computing of quantum chemistry. First, GPUs typically use high-speed, but limited-capacity memory which is significantly smaller than sysmtem RAM on a CPU-based system. The large matrices and complex wave functions can quickly exceed GPU memory limits. It necessitates either offloading some of the data back to the CPU or employing sophisticated memory management techniques, both of which can degrade performance. Second, the bandwidth between CPU memory and GPU memory is limited and can become a significant bottleneck, especially in applications where the GPU needs to frequently access or update data stored in main system memory. To prevent the data transfer, a large amounts of classic CPU routines have to be reimplemented on GPU. Third, Maximizing GPU performance often involves fine-tuning the code based on GPU architecture, such as memory access patterns, thread management, and kernel configurations, which can be a highly specialized and time-intensive process.

Optimizing a quantum chemistry workflow that balances chemical accuracy and computational efficiency requires extensive domain expertise and practical experience. Initially, a quantum chemist is tasked with selecting an appropriate basis set and exchange-correlation functional from a broad dataset, exemplified by the Basis Set Exchange, which catalogs over 600 basis sets~\cite{bse}, and LibXC, comprising more than 400 exchange-correlation functionals~\cite{libxc}. Subsequent considerations include evaluating methods for dispersion correction, solvent model selection, basis set superposition error (BSSE) correction, and establishing convergence criteria for various computational tasks. The adoption of recent composite DFT protocols, such as PBEh-3c~\cite{pbeh-3c}, $\omega$B97X-3c~\cite{wb97x-3c}, and r2SCAN-3c~\cite{r2scan-3c}, necessitates precise configurations of basis sets, exchange-correlation functionals, BSSE corrections, and dispersion corrections. Additionally, contemporary quantum chemistry workflows frequently incorporate integration with other software packages, including RDKit, PyTorch, and Qiskit, to extend their computational capabilities. Notably, the open-source communities behind PySCF~\cite{pyscf} and Psi4~\cite{psi4} have contributed user-friendly Python interfaces, offering a wide range of functionalities. PySCF, in particular, is recognized for its straightforward, Pythonic interface, facilitating seamless integration with other Python libraries and enabling the scripting of customized quantum chemistry workflows. Its adaptable design also supports acceleration within the modern GPU ecosystem, underscoring the importance of flexible, integrated approaches in the development of efficient quantum chemistry workflows.

In this work, we will present our contributions to the GPU4PySCF package and related modules to enhance its functionality. GPU4PySCF is an open-source project that was initiated in the Chan group \cite{gpu4pyscf-caltech} to add GPU acceleration to key parts of the PySCF computational workflow. At the time that we started our work, GPU4PySCF contained CUDA kernels for 4-center two-electron integrals and GPU accelerated direct SCF (Hartree-Fock) functionality \cite{gpu4pyscf-caltech}. Our contributions enhance the capabilities of this module in several key areas: 1) through density fitting algorithms. These adapt the Rys quadrature CUDA kernels in GPU4PySCF to the 3-center quantities for density fitting. However, once the density fitted integrals are computed, the remaining operations are tensor operations, which are inherently more GPU-friendly. The density fitting implementation is optimized for small to medium sized molecules, and thus complements the existing functionality for direct Self-Consistent Field (SCF) methods that can be extended to larger molecules. 2) Optimizing our codebase to leverage modern GPU architectures, such as tensor cores, to achieve up to twice the speed in tensor contractions. 3) Integrating a wide array of open-source packages from the quantum chemistry community into the PySCF package, thus supporting sophisticated DFT methods, popular basis sets, solvent models, chemical property calculations, geometry optimization, and transition state search methodologies. Through these comprehensive contributions, we can now consider the GPU4PySCF package to be feature-rich and an industrially relevant quantum chemistry tool.

%In this work, a Python-based, and GPU-based framework is built on the top of PySCF. Technically, we push the evolop of quantum chemistry calculations on GPUs in several directions. 1) We focus on the density fitting algorithms which is more friendly to GPUs. Direct SCF is supported with limited functionalities. 2) At linear algebra level, we adapted the code to the modern GPU architecture such as tensor cores, which deliver as much as 2x faster tensor contractions. 3) Fruitful open-sourced packages in the quantum chemistry community are incorporated with the package GPU4PySCF. Sophisticated DFT methods, widely-used basis sets, solvent models, chemical properties, geometry optimization, and transition state search are supported.

With our contributions, we underscore the notable advantages of GPU4PySCF as follows:
\begin{itemize}
    \item {\bf Speed}: GPU4PySCF now provides performance that is equivalent to
    that of 600-1000 CPU cores for running conventional CPU-based quantum chemistry software.
    As a result, it offers substantial cost savings, for instance, up to 90\% when
    utilizing the NVIDIA A100-80G cards.
    \item {\bf Python Integration}: The framework allows direct access to quantum features of atoms at the Python level, ensuring seamless compatibility with other Python-based packages such as PyTorch, Jax, TensorFlow, RDkit, and more.
    \item {\bf Open Source}: Our contributions to GPU4PySCF benefits from the support of the PySCF community, the largest open-source quantum chemistry community, fostering collaborative development and innovation.
    \item {\bf Application Driven}: With our contributions that are specifically optimized and tailored for industrial applications, GPU4PySCF excels in tasks such as DFT calculations, offering practical advantages for commercial research and development.
\end{itemize}

Detailed performance metrics and benchmarks are provided in Section \ref{sec:benchmark}. These calculations have been rigorously cross-validated with Q-Chem 6.1, ensuring reliability and accuracy. Furthermore, in Section \ref{sec:applications} briefly illustrates how GPU4PySCF v1.0 can be utilized to navigate the potential energy landscape, analyze quantum features, compute thermodynamic properties, estimate solvation free energies, quantify chemical reactions, and integrate with neural network models. This comprehensive approach showcases the framework's versatility and potential to revolutionize quantum chemistry computations through GPU acceleration.

%We highlight the following advantages of GPU4PySCF
%\begin{itemize}
%    \item Fast. 600-1000x faster than typical CPU-based quantum chemistry softwares on a sigle core, and saves as much as 90\% of the actual cost with NVIDIA A100-80G.
%    \item Python-based. Quantum features of atoms are accessible at Python level. It is compatible with other python-based packages such as PyTorch, Jax, TensorFlow, RDkit and so on.
%    \item Open-sourced. Its development is supported by PySCF community, which is largest open-sourced quantum chemistry community.
%    \item GPU4PySCF is optimized and prioritized for industrial applications, such as DFT calculations.
%\end{itemize}

%The performance and benchmarks are given in Section \ref{sec:benchmark}. The calculations are cross validated with Q-Chem 6.1. In Section \ref{sec:applications}, we briefly showcase that how GPU4PySCF can be used to explore the potential energy surface, analyze quantum features, evaluate thermodynamic properties, estimate solvation free energy, quantitize chemical reactions, and incorporate with neural network models.

\section{Performance and Benchmark}
\label{sec:benchmark}
\subsection{Dataset and Performance}
We focus on benchmarks with the density fitting scheme, using the optimized implementation that we have contributed. This is efficient for small molecules with less than 200 atoms. A small and diverse dataset is constructed for the benchmarking the fundamental modules in GPU4PySCF. We select small molecules ($<$ 100 atoms) from the supplement information in~\cite{dataset} and two medium-sized molecules from the supplement information in~\cite{brainqc2}. The newly constructed dataset includes 8 elements (H, C, O, N, Mg, S, Cl, P), and 13 small molecules with 20-168 atoms. The details of the dataset are provided in Appendix~\ref{sec:dataset}. The xyz files and benchmarking details can be found on \href{https://github.com/pyscf/gpu4pyscf/tree/master/benchmarks}{GitHub}. Other transition metals, such as Nb, Ti, Ru and so on, are also supported in GPU4PySCF, although they are not included in this dataset. This capability will be shown in Sec.~\ref{sec:ts}. Larger molecules with more than 168 atoms can be calculated within the existing direct SCF scheme in GPU4PySCF~\cite{gpu4pyscf-caltech}, which we do not focus on here.

\begin{figure}[!htp]
    \centering
    \includegraphics[width=0.7\linewidth]{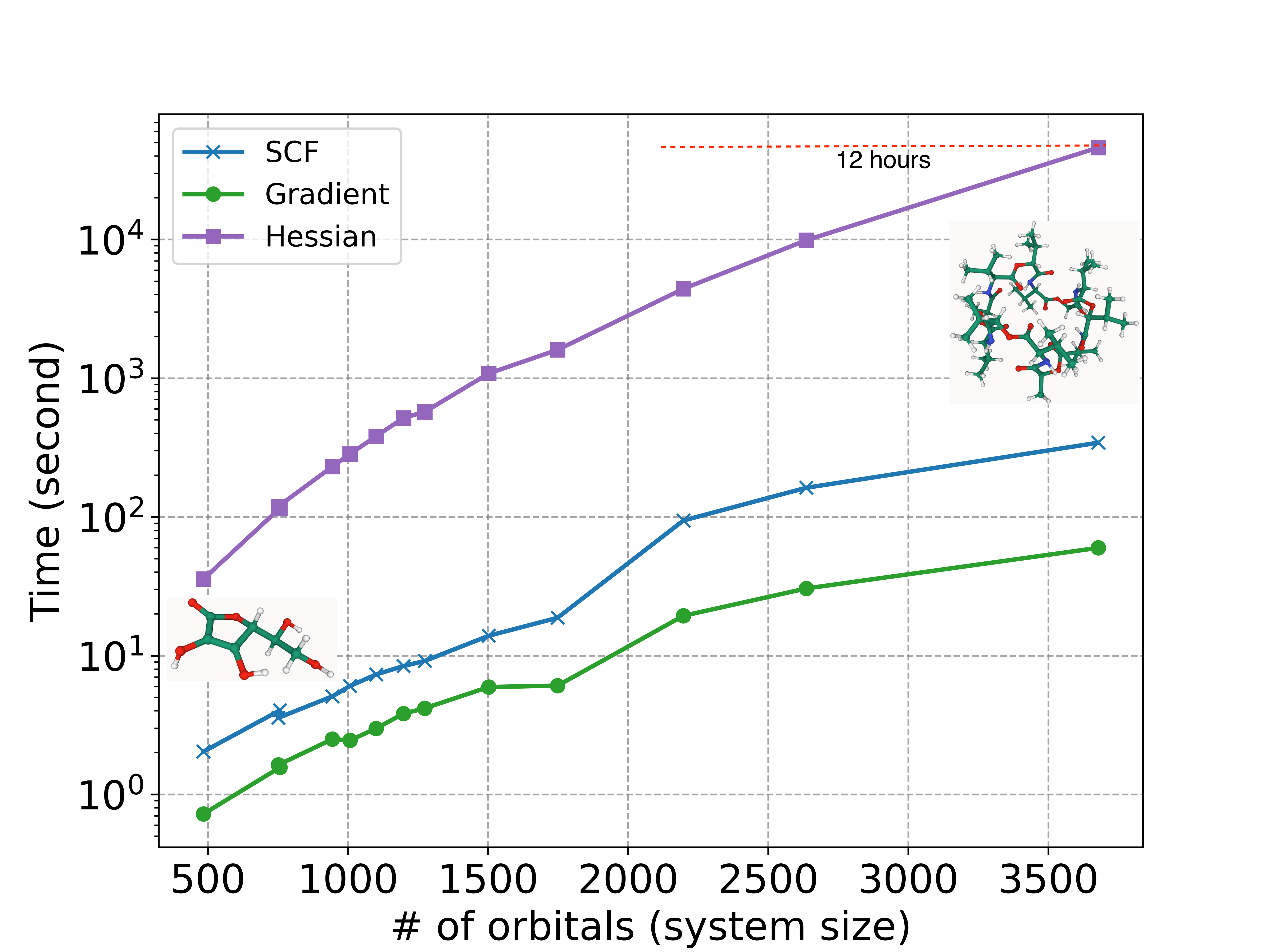}
    \caption{Scaling curve of SCF, gradient, and Hessian calculations. def2-TZVPP basis set, def2-universal-JKFIT auxiliary basis, B3LYP, (99,590) grids, on A100-80G.}
    \label{fig:scaling}
\end{figure}

For a typical DFT protocol with def2-TZVPP, B3LYP, and (99,590) grids, we record the wall clock time for SCF, gradient, and Hessian calculations on NVIDIA A100-80G. For the smallest molecule (Vitamin C) in the dataset, SCF, gradient, and Hessian take 1.7, 0.68, and 43 seconds respectively. For the largest molecule (Valinomycin) in the dataset, SCF and gradient calculations take 309 seconds and 58 seconds respectively. The Hessian calculation of Sphingomyelin (84 atoms) takes 30 minutes. Hessian calculation is the most expensive module of the vibrational analysis. The vibrational analysis of moderately sized (100-200 atoms) molecules is prohibitively expensive with traditional quantum chemistry softwares~\cite{xtb}. Unfortunately, coupled-perturbed SCF iterations for (Azadirachtin, Taxol, Valinomycin) does not converge. The time scaling curve of SCF, gradient, and Hessian calculations is shown in Figure~\ref{fig:scaling}. Theoretically, SCF and gradient calculations are quartic scaling with respect to the system size, while Hessian calculation is $O(N^5)$. The slope of the scaling curve decreases as the system size gets larger, due to sparsity. The bottleneck of calculating a large molecule is the CPU memory storage. A CPU-GPU hybrid strategy is employed to store GTO integrals in this work. For small molecules, all the intermediate variables are stored in GPU memory. For the large system, the variables are stored in CPU memory. The data in CPU memory is asynchronously transferred to GPU memory for consumption.

\subsection {Speedup of SCF, Gradient, and Hessian Calculations}
Achieving complete alignment between various quantum chemistry software settings presents a significant challenge. Fundamental configurations like the threshold for two-electron integrals, exchange-correlation functionals, basis sets, and Lebedev grids are generally standardized within the quantum chemistry community. However, more intricate adjustments—such as mitigating the linear dependence of atomic orbitals, defining SCF convergence criteria, and pruning DFT grids—often exhibit subtle discrepancies. Moreover, benchmarking these results against commercial software treated as a "black box" adds an additional layer of complexity. In our analysis, we strive to minimize these discrepancies by uniformly applying stringent criteria across basic settings. Given the restricted access to commercial quantum chemistry software, our benchmarking efforts are focused exclusively on Q-Chem, a prominent commercial platform renowned for its cutting-edge algorithms designed for large systems. The discrepancies between Q-Chem 6.1 and GPU4PySCF for exchange-correlation functionals and different basis sets are shown in Appendix \ref{sec:accuracy_xc} and \ref{sec:accuracy_basis}.

Our benchmarking endeavors concentrate on three foundational modules: Self-Consistent Field (SCF), gradient, and Hessian calculations. These modules are the most expensive components in numerous quantum chemistry computations. In this section, all computations are conducted on closed-shell systems using density fitting. The functionalities of open-shell calculations will be shown in Sec. \ref{sec:redox} and Sec. \ref{sec:fukui}. Notably, due to the memory management issue within the density fitting coupled perturbed SCF iteration in Q-Chem, the reference Hessian calculations are performed using finite difference methods. But the Hessian matrices are calculated analytically in GPU4PySCF. We anticipate that resolving these issues could enhance the efficiency of Hessian calculations in Q-Chem by a factor of 2-3. Hessian calculations for molecules exceeding 84 atoms become prohibitively expensive for traditional quantum chemistry packages. We only benchmark the speedup across molecules containing 20-84 atoms.

% It is a challenging task to completely align the settings of different quantum chemistry softwares. Most basic settings such as threshold of two-electron integrals, exchange-correlation functionals, basis set and Lebedev grids are quite standard across the quantum chemistry community. However, more sophisticated settings, such as removing linear dependence of atomic orbitals, SCF convergence criteria, and pruning DFT grids,  can be subtly different. It is even more challenging to benchmark the results with the commercial softwares as a blackbox. In this article, we use the same basic settings, and reduce the difference between two softwares by tightening all criteria. With the limited access of the commercial quantum chemistry softwares, we only benchmark our results with Q-Chem. Q-Chem is a well-established commercial quantum chemistry software with innovative algorithms for large system.

% We focus on benchmarking three basic modules SCF, gradient, and Hessian. Other quantum chemistry calculations often require one or more of those three modules. The additional cost of other calculations are often not expensive. All the calculations are closed-shell, and performed with density fitting. Due to the memory management issue in density fitting coupled perturbed SCF, Hessian calculations are calculated with finite difference in Q-Chem. We expect the performance of Hessian calculation in Q-Chem to be 2-3x faster if the issue is fixed. The speedup is averaged over molecules with 20-84 atoms. The Hessian calculations for more than 84 atoms are prohibitively expensive.

\begin{figure}[!htp]
    \centering
    \includegraphics[width=0.9\linewidth]{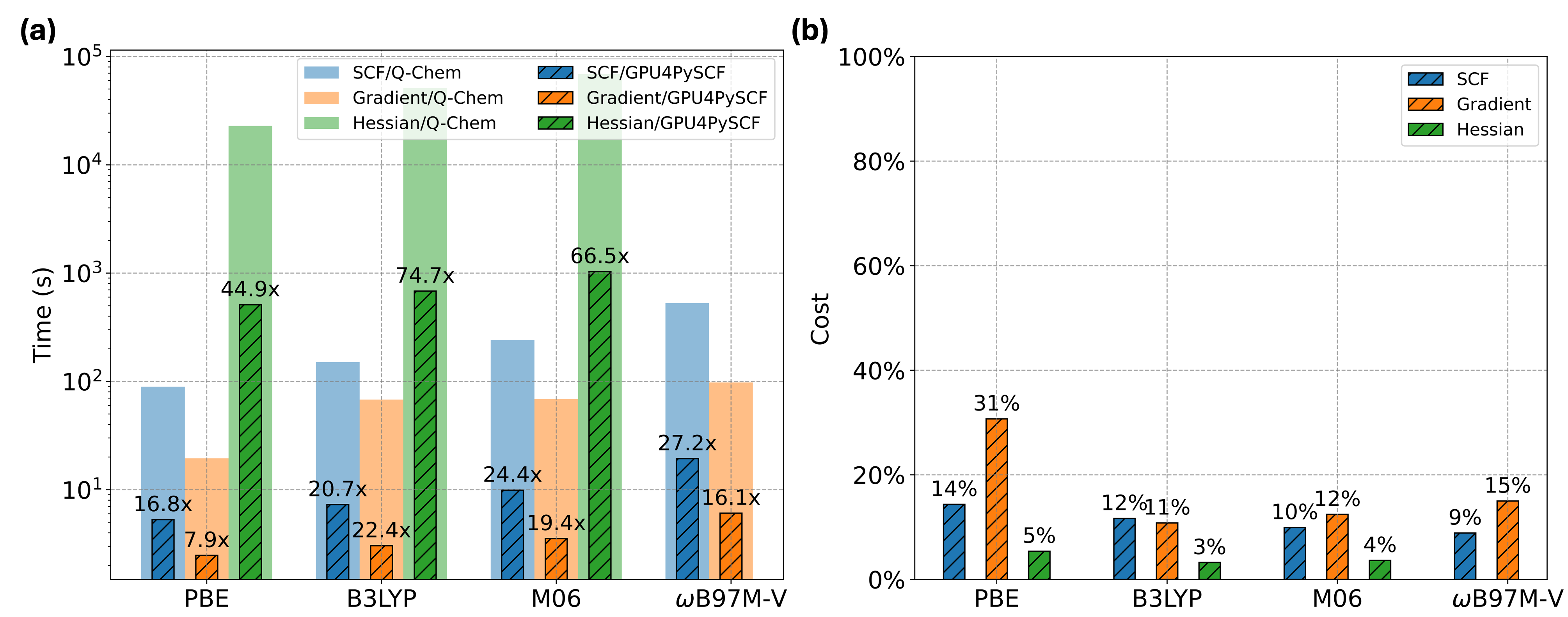}
    \caption{Speedup and actual cost saving of GPU4PySCF on A100-80G over Q-Chem v6.1 on 32-core CPU. The cost is estimated based on AWS pricing. \$40.966/hr for {\bf p4de.24xlarge} GPU instance with 8 A100-80G GPUs, 96 vCPUs and 1152 GiB CPU memory. \$2.117/hr for {\bf r7i.8xlarge} CPU instance with 32 vCPUs and 256 GiB memory. def2-TZVPP basis set, def2-universal-JKFIT auxiliary basis, (99,590) XC grids, (50,194) NLC grids.}
    \label{fig:speedup}
\end{figure}
On average, GPU4PySCF achieves a performance that is 20 times faster than identical SCF calculations performed using Q-Chem 6.1 on 32 CPU cores. This speedup increases dramatically to 50 times for Hessian calculations, although the acceleration observed in gradient calculations is relatively modest. However, since gradient calculations are generally quicker than SCF iterations, this does not significantly affect the overall efficiency of geometry optimization or transition state search tasks. An evaluation of the actual costs based on AWS pricing for comparable machinery shows that GPU4PySCF can reduce expenses by about 90\% across most tasks. For the gradient calculation with pure DFT, the saving decreases to 70\%. This task technically can be accelerated with a more efficient algorithm~\cite{j-engine}. We leave this implementation as future work. It is observed that larger molecules typically experience a more substantial speedup compared to smaller molecules. This is due to the high GPU occupancy of a large molecule.

\subsection{Speedup of DFT with Implicit Solvent Models}
Solvent models in quantum chemistry critical for simulating and understanding the behavior of molecules in solution, an essential aspect given most chemical reactions occur in some type of solvent. The Polarizable Continuum Model (PCM) models are the most common implicit solvent models. But the computational efficiency of those models are rarely discussed. Especially when the density fitting scheme is employed, the computational cost of PCM models is not negligible. Since Q-Chem 6.1 does not support density fitting SCF with PCM models, we use the standard SCF scheme as a reference. Thus, the following speedups should be interpreted as two factors: GPU speedup and density fitting speedup. We stress that the different algorithms are used in Q-Chem and GPU4PySCF. The comparison is performed to mimic the actual usage of the solvent models. The discrepancies between Q-Chem 6.1 and GPU4PySCF for solvent models are shown in Appendix \ref{sec:accuracy_solvent}.

We benchmark DFT calculations employing two PCM models, C-PCM~\cite{cpcm1,cpcm2} and IEF-PCM~\cite{iefpcm1}. IEF-PCM model is slightly more expensive than C-PCM model since more entries are calculated for the linear system. The SMD model~\cite{SMD} introduces the additional computation of CDS contributions on the top of IEF-PCM model. Yet, the computational cost of these contributions is almost negligible. The efficiency of CDS contributions will not be discussed in this section, but the accuracy benchmarks will be presented in Section \ref{sec:solvation_free_energy}. Other types of implicit solvent models, such as generalized Born models, are not implemented in the current version.

We evaluate the performance of molecules with up to 42 atoms. The computational cost of larger molecules is prohibitively expensive with Q-Chem, although GPU4PySCF is able to handle large molecules. With the A100-80G GPU and density fitting, GPU4PySCF significantly accelerates the SCF calculations by 40-80 times, gradient calculations by 20-40 times, and Hessian calculations by 100-170 times, respectively. The speedup of the calculations in the gas phase is slightly lower than the results in Table \ref{tab:solvent}. For instance, the speedup of SCF for Vitamin C in gas phase is around 33. The speedup difference between gas phase and liquid phase suggests the high efficiency of PCM models in GPU4PySCF.

%On average, GPU4PySCF performs 20x faster than the same SCF calculations with Q-Chem 6.1 on 32-CPU cores. The speedup boosts up to 50x for the Hessian calculations. The speedup of gradient calculation is relatively low. However, it is usually much faster than SCF iterations. It does not affect the overall speedup of geometry optimization or transition state search tasks too much. We estimated the actual cost of AWS pricing for the similar machines. GPU4PySCF saves around 90\% of the actual cost for most tasks. The savings drops to 70\% for gradient calculation of pure DFT. The large molecule generally has a higher speedup than the small molecules (SI).

%We benchmark DFT calculations with two PCM models, C-PCM~\cite{cpcm1,cpcm2} and IEF-PCM~\cite{iefpcm1}. SMD model requires the additional computaion of CDS contributions. But the computational cost of CSD contributions is almost neglectible. We will not show the its efficiency in this section. The accuracy benchmarks will be presented in \ref{sec:solvation_free_energy} Since density fitting for PCM models is not supported yet in Q-Chem 6.1, the standard SCF scheme is used as the reference. Due to the expensive computational cost, we only perform the calculations for molecules with up to 42 atoms with Q-Chem. With A100-80G GPU and density fitting, GPU4PySCF accelerates SCF calculations for 40-80 times, gradient calculations for 20-40 times, and Hessian calculations for 100-170 times respecively.

\begin{table}[!htp]
    \centering
    \begin{tabular}{|c|ccc|ccc|}
    \hline
        Molecule               &   \multicolumn{3}{|c|}{C-PCM}  & \multicolumn{3}{|c|}{IEF-PCM} \\
                           &    SCF &  Gradient  &    Hessian &      SCF &     Gradient &   Hessian \\
    \hline
        Vitamin C     & 39.8 &  21.6 & 107.2 &   31.0 &   18.3 &  100.1 \\
        Inosine       & 64.3 &  30.4 & 162.3 &   54.3 &   30.6 &  158.2 \\
        Bisphenol A   & 72.8 &  35.9 & 158.7 &   67.1 &   37.6 &  155.2 \\
        Mg Porphin    & 86.8 &  40.9 & 168.5 &   81.1 &   41.4 &  166.4 \\
        Penicillin V  & 78.0 &  36.8 & 186.0 &   81.0 &   36.7 &  186.5 \\
    \hline
    \end{tabular}
    \caption[short]{Speedup of GPU4PySCF on A100-80G over Q-Chem 6.1 on 32-core CPUs for DFT tasks with PCM solvent models. B3LYP, def2-TZVPP, def2-universal-JKFIT, (99,590), 302 Lebedev grids for both H atoms and Heavy atoms.}
    \label{tab:solvent}
\end{table}

\section{Applications}
\label{sec:applications}
In this section, we take the opportunity to showcase the capabilities of of GPU4PySCF v1.0 with our new contributions in several industrial applications. A complete list of capabilities will be shown in Section \ref{sec:functionalities}.
\subsection{Exploring Potential Energy Surface}
\subsubsection{Torsion Scan}
The development of molecular mechanics force fields significantly benefits from torsion scans conducted at precise quantum chemistry levels, serving as a crucial source of training data~\cite{Boothroyd_Behara_Madin_Hahn_Jang_Gapsys_Wagner_Horton_Dotson_Thompson_et,Horton_Boothroyd_Wagner_Mitchell_Gokey_Dotson_Behara_Ramaswamy_Mackey_Chodera_et,Qiu_Smith_Stern_Feng_Jang_Wang_2020,Qiu_Smith_Boothroyd_Jang_Hahn_Wagner_Bannan_Gokey_Lim_Stern_et,Roos_Wu_Damm_Reboul_Stevenson_Lu_Dahlgren_Mondal_Chen_Wang,Wang_Wolf_Caldwell_Kollman_Case_2004,Xue_Yang_Zhang_Wan_Fang_Lin_Sun_Gobbo_Cao_Mathiowetz}. However, torsion scans are notably time-consuming, as they require at least 24 independent constrained geometry optimizations for each proper torsion angle in a molecule. These optimizations determine the potential energy of the molecule with the torsion angle fixed at specific values within the $[-180^{\circ}, +180^{\circ}]$ range. Depending on the complexity of the potential energy surface, each optimization might necessitate 10-100 sequential quantum chemistry calculations, underscoring the importance of computational efficiency for this task. Historically, such scans have been restricted to small molecule fragments, typically using small basis sets at the double-zeta level.

The advent of GPU4PySCF, coupled with its seamless integration with the Pythonic interface of the geometric optimizer geomeTRIC, revolutionizes this process, allowing for torsion scans on a scale previously deemed unfeasible. This enables routine computations of torsion scans for real-world drug molecules employing large basis sets.

\begin{figure}[!htp]
    \centering
    \includegraphics[width=0.9\linewidth]{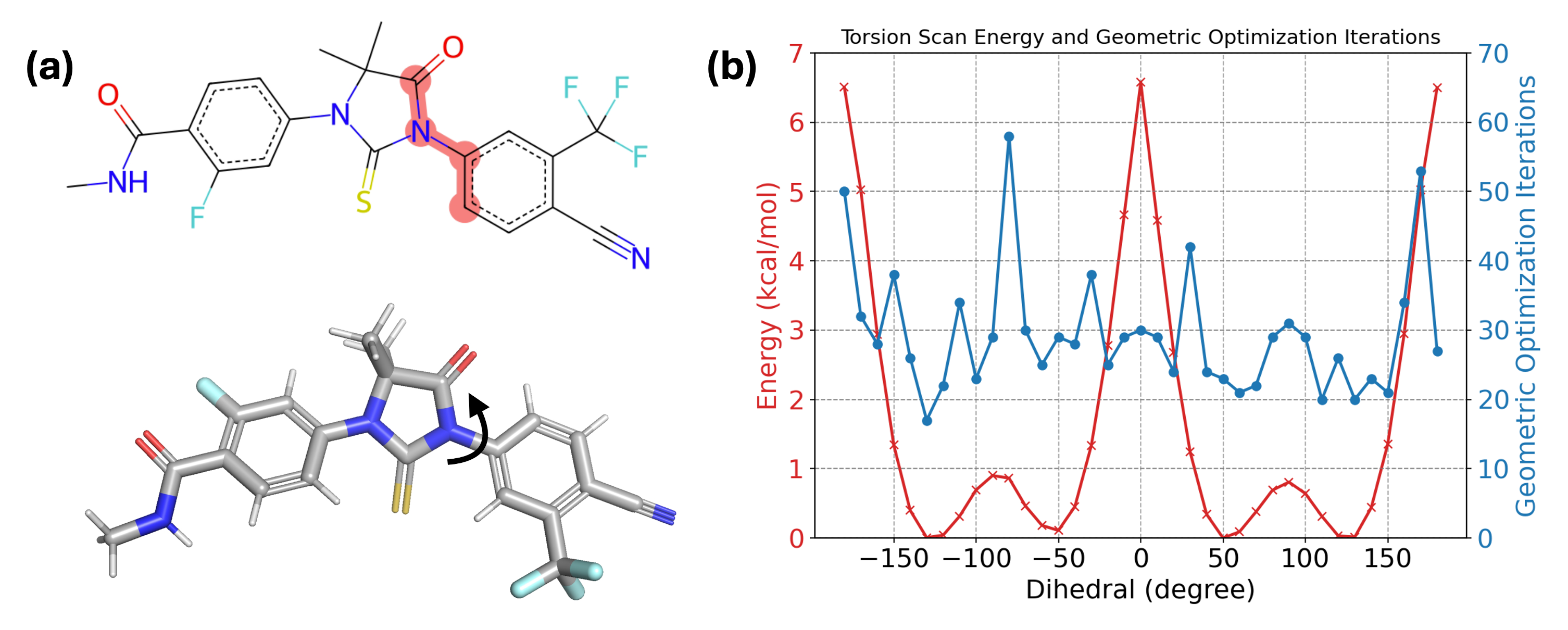}
    \caption{
    \label{fig:torsionscan}
    An example of torsion scan.
    (a) The molecule Enzalutamide (PubChem CID 15951529).
    (b) Torsion scan energy profile and number of geometry optimization iterations (number of DFT calculations) at each scanned proper torsion angle degree.
    }
\end{figure}

An illustrative example of this capability is provided by our work on Enzalutamide (Figure~\ref{fig:torsionscan}). Utilizing GPU4PySCF, we conducted a torsional energy profile scan using $\omega$B97M-V/def2-TZVPP at $10^{\circ}$ step (37 angles in total). This comprehensive analysis entailed 1089 DFT evaluations, with the entire scan completed in 34,679.9 seconds, averaging approximately 32 seconds per DFT evaluation. This ability to scan large, complex molecules using high-precision quantum chemistry algorithms significantly broadens the scope of chemical spaces accessible through torsion scan methodologies. All calculations in this subsection were performed using grid level 3 in PySCF with density fitting on an A100-80G GPU.

\subsubsection{Dimer Interaction Energies}
Noncovalent interactions are paramount in molecular modeling, especially for the accurate prediction of thermodynamics and transport properties~\cite{jorgensen1988opls,sun1998compass}. Additionally, robust noncovalent interactions, such as hydrogen bonding and $\pi$-$\pi$ stacking, are essential for ligand-protein binding~\cite{abraham2002application,bootsma2019predicting,Roos_Wu_Damm_Reboul_Stevenson_Lu_Dahlgren_Mondal_Chen_Wang,lu2021opls4}. The assessment of dimer interaction energies through quantum mechanical approaches serves as a prevalent method to explore noncovalent interactions. However, attaining precise noncovalent interaction energies demands high-fidelity techniques, which invariably elevate computational costs. Thanks to the efficiency of GPU4PySCF, interaction energies for systems comprising hundreds of atoms are now attainable within practical timeframes, even when utilizing sophisticated DFT methodologies that incorporate non-local correlation effects.

\begin{figure}[!htp]
    \centering
    \includegraphics[width=0.9\linewidth]{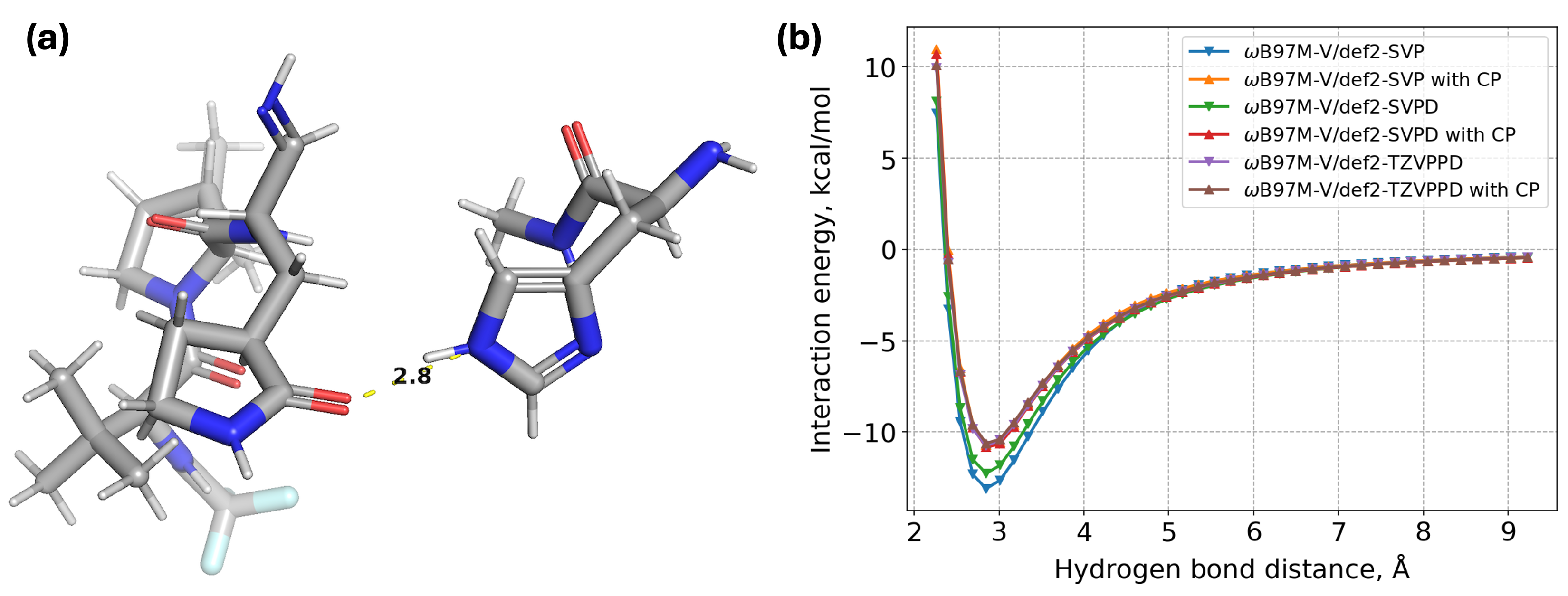}
    \caption{
    \label{fig:int_energy}
    An example of noncovalent interaction.
    (a) Nirmatrelvir (left) and the histidine residue (right).
    The hydrogen bond distance (between donor and acceptor atoms) is 2.8 \r{A} in the shown geometry, which is truncated from PDB 7yrz.
    (b) Interaction energy at varying hydrogen bond distances.
    }
\end{figure}

We offer an illustrative example involving the oral COVID-19 antiviral Nirmatrelvir~\cite{halford2022path} (PubChem CID 155903259, Figure~\ref{fig:int_energy} (a)), a SARS-CoV-2 main protease inhibitor developed by Pfizer. In its binding pocket, Nirmatrelvir forms a hydrogen bond with a histidine side chain. The interaction energy between Nirmatrelvir and the histidine residue (Figure~\ref{fig:int_energy} (b)) is calculated using various bases. For the direct evaluation of interaction energy, 42 single-point energy calculations were conducted. The total wall times for def2-SVP, def2-SVPD, and def2-TZVPPD bases were 2370, 3082, and 5823 seconds, respectively. Moreover, the counterpoise method (CP)~\cite{van1994state} was applied to mitigate basis set superposition error (BSSE)~\cite{balabin2008enthalpy,hobza2010non}, involving 120 single-point energy calculations for each base, with total wall times of 4020, 5510, and 11892 seconds, respectively. All these calculations were performed using the $\omega$B97M-V functional and grid level 3 in PySCF with density fitting on an A100-80G GPU.

\subsubsection{Lithium Ion Solvation Structure}
\label{subsec:battery_cluster}
%Studying the solvation structure of $\mathrm{Li^+}$ ions is crucial for developing $\mathrm{Li^+}$ ion battery with high energy density, rapid charging, low cost, high sustainability, as it guides the design of electrolytes\cite{meng2022designing}. 

%Solvation structures play a crucial role for the advancement of liquid electrolyte designs \cite{meng2022designing}. 
The concept of solvation structure engineering has been utilized in recent studies, such as high-concentrated \cite{wang2016superconcentrated}, localized high-concentrated \cite{chen2018high}, and fluorinated ether-based electrolytes \cite{yu2020molecular,amanchukwu2020new}. By engineering solvation structures, these studies 
show improvements in ion transport, solid electrolyte interphase, and electrochemical stability \cite{borodin2017liquid,xu2007solvation,borodin2020uncharted, yamada2019advances}.
%which lead to the realization of high energy density, fast charging, and cell lifetime. 
%Recent electrolyte designs, such as high-concentrated \cite{wang2016superconcentrated}, localized high-concentrated \cite{chen2018high}, and fluorinated ether-based electrolytes \cite{yu2020molecular,amanchukwu2020new}, leverage solvation structure engineering to enable efficient ion transport, stable solid electrolyte interphase, and improved electrochemical stability \cite{borodin2017liquid,xu2007solvation,borodin2020uncharted, yamada2019advances}. These improvements facilitate the realization of high energy density, fast charging, and cell lifetime. 
Binding energy of solvation structures is an indicator of solvents' ability to dissolve salt \cite{chen2019cation}. In addition, binding energy plays a crucial role in solvation and desolvation, which is the mechanism for Li$^+$ plating/stripping and intercalation in Li batteries \cite{zhang2021electrolyte, moon2014mechanism}. As such, investigating solvation structures and their corresponding binding energies facilitate the realization of high energy density, fast charging and improved cell lifetime. 

\begin{figure}[htp]
    \centering
    \includegraphics[width=1.0\linewidth]{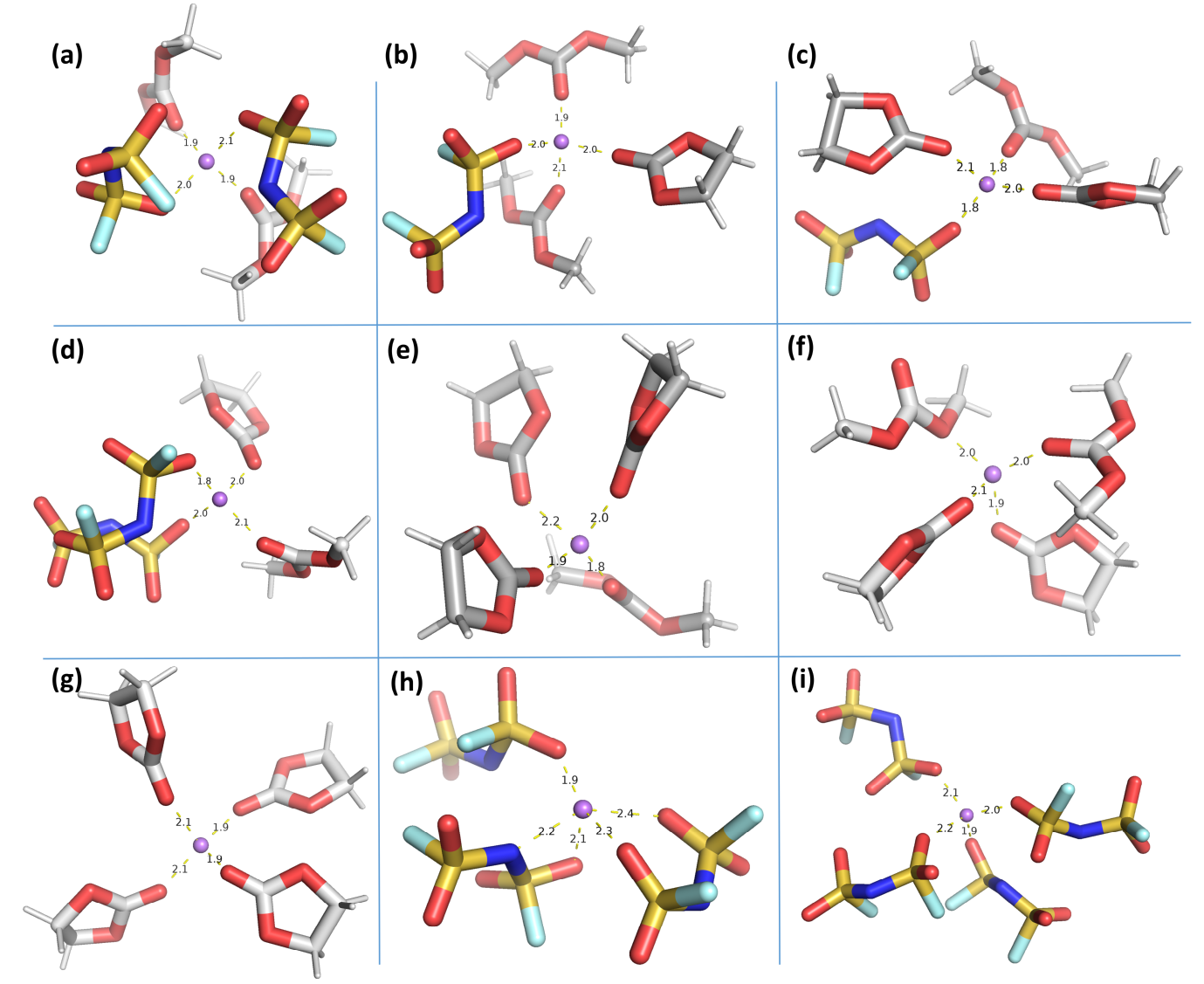}
    \caption{Solvation structures of $\mathrm{Li^+}$ ion. The clusters depicted in images (a) to (i) correspond respectively to the clusters listed in Table \ref{tab:li_bindingenergy}. The distances between $\mathrm{Li^+}$ ion and coordinated atoms are plotted in \AA.}
    \label{fig:battery_total}
\end{figure}

%For a simple demonstration, we correlate the binding energy of solvation structures with the respective solvents' capacity to dissolve salt.
In this work, we examine the solvation structures of $\mathrm{Li^+}$ ions in the electrolyte by computing the structure and binding energy of $\mathrm{Li^+}$ ion clusters. Through molecular dynamics simulations of the battery electrolyte containing 2.25M LiFSI(lithium bis(fluorosulfonyl)imide) in DMC(dimethyl carbonate):EC(ethylene carbonate) with a weight percentage ratio of 51:49, we captured typical solvation structures of $\mathrm{Li^+}$ ions and computed the binding energies of these structures using GPU4PySCF, as presented in Figure \ref{fig:battery_total} and Table \ref{tab:li_bindingenergy}.

\begin{figure}[htp]
    \centering
    \includegraphics[width=0.8\linewidth]{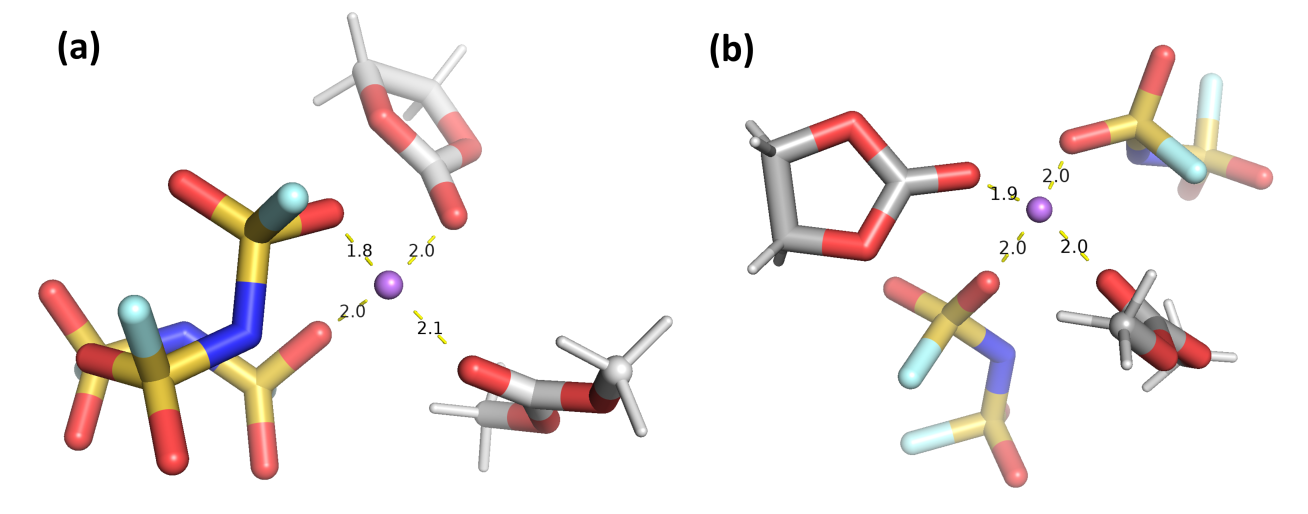}
    \caption{Two different structures of  \( \text{Li}^+(\text{DMC})(\text{EC})(\text{FSI}^-)_2 \). The energies of clusters in (a) and (b) are -1.078 eV and -0.941 eV, respecively. The distances between $\mathrm{Li^+}$ ion and coordinated atoms are plotted in \AA.}
    \label{fig:battery_compare}
\end{figure}

Most of the clusters we examined have a coordination number of 4, consisting of either solvent or anion molecules, or a combination of the two. Their binding energies range from -0.065 eV to -1.078 eV, indicating significant differences in the energy stability of the clusters. 
We observe that clusters composed solely of Li$^+$ and FSI$^-$ exhibit the most positive binding energies compared to those containing solvents. This observation aligns with chemical intuition, suggesting that the salt dissolution of LiFSI in an EC-DMC polar solvent mixture is preferred, manifesting the fundamental requirement of solvent functionality in Li-battery liquid electrolyte applications. 
Besides, clusters with the same composition can exhibit different energies if their structures vary, as shown in Figure \ref{fig:battery_compare} for the \( \text{Li}^+(\text{DMC})(\text{EC})(\text{FSI}^-)_2 \) cluster, suggesting adequate geometric sampling for DFT calculations is necessary for accurate representation of solvation structure stabilities, emphasizing the importance of computational efficiency provided in the code. 

%Based on our analysis, we can preliminarily conclude that the most prevalent solvation structure of $\mathrm{Li^+}$ ions in the electrolyte is a tetra-coordinated structure combining both anions and solvent molecules. 

All calculations are performed with 6-311++G(3df,3pd) basis set and cc-pVTZ-RI auxiliary basis for the density fitting technique. The IEF-PCM with UFF radii was utilized for modeling solvent effects, with the dielectric constant being 20.0. Additionally, M06-2X functional together with Grimme's D3 version of dispersion correction with zero damping was used. Note that our examples in this section are merely illustrative and do not provide a complete study of the issue of $\mathrm{Li^+}$ ion solvation structures.

\begin{table}[ht]
    \centering
    \begin{tabular}{ll}
    \hline
    Cluster & Binding Energy \\
    \hline
    \( \text{Li}^+(\text{DMC})_2(\text{FSI}^-)_2 \) & -1.077 \\
    \( \text{Li}^+(\text{DMC})_2(\text{EC})(\text{FSI}^-) \) & -1.020 \\
    \( \text{Li}^+(\text{DMC})(\text{EC})_2(\text{FSI}^-) \) & -1.018 \\
    \( \text{Li}^+(\text{DMC})(\text{EC})(\text{FSI}^-)_2 \) & -1.004 \\
    \( \text{Li}^+(\text{DMC})(\text{EC})_3 \) & -0.893 \\
    \( \text{Li}^+(\text{DMC})_2(\text{EC})_2 \) & -0.857 \\
    \( \text{Li}^+(\text{EC})_4 \) & -0.735 \\
    \( \text{Li}^+(\text{FSI}^-)_3 \) & -0.656 \\
    \( \text{Li}^+(\text{FSI}^-)_4 \) & -0.065 \\
    \hline
    \end{tabular}
    \caption{Binding energies (eV) of various clusters.}
    \label{tab:li_bindingenergy}
\end{table} 

\subsection{Analyzing Molecular Properties}
\subsubsection{CHELPG Charge Calculation}
CHELPG (charges from electrostatic potentials using a grid-based method) is a computational method designed to calculate atomic charges within a molecule based on the electrostatic potential generated by its electrons~\cite{breneman1990determining,manz2020seven}. Introduced by Breneman and Wiberg in 1990, this method aims to derive atomic charges that are reflective of the molecule's electrostatic potential on a grid surrounding the molecule~\cite{breneman1990determining}. These charges can be then utilized in simulations to model intermolecular forces and other electrostatic properties. The essence and the most time-consuming part of CHELPG lies in calculating the electrostatic potential at grids around the molecule. The method then optimizes the atomic charges by fitting these calculated potentials to the actual electrostatic potential observed, using a least-squares method. This optimization process ensures that the derived atomic charges closely represent the true electrostatic potential around the molecule.

\begin{figure}[!htp]
    \centering
    \includegraphics[width=0.8\linewidth]{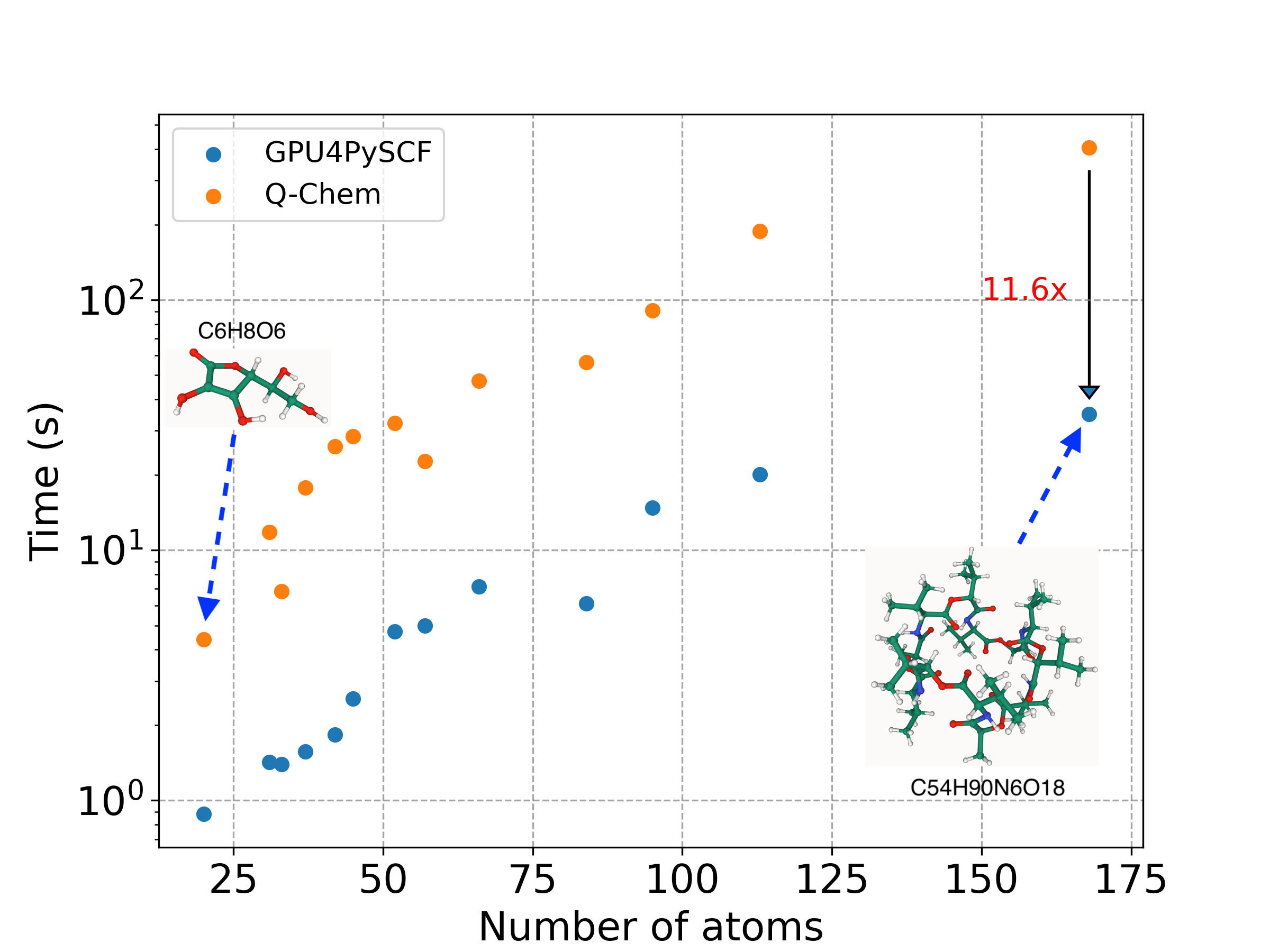}
    \caption{Comparison of the time required for CHELPG calculation between GPU4PySCF and Q-Chem. The computations by Q-Chem used 64 cores and 128G memory, while GPU4PySCF used 15 cores, 245G memory, and one A100-SXM-80GB GPU. The tested systems are benchmark molecules listed in Table \ref{tab:dataset}.
    \label{fig:chelpgtime}
    }
\end{figure}

Followed by the work of Herbert and co-workers~\cite{herbert2012rapid}, CHELPG is implemented in the package, and the computational efficiency has been enhanced with the aid of GPUs. Our implementation uses the Cartesian grid with the default spacing 0.3 \AA~and the radial extent of the CHELPG grid 2.8 \AA~as recommended in Ref.~\cite{breneman1990determining}. Besides, the smoothing function used in Ref.~\cite{herbert2012rapid} is also implemented, where the Bondi radius is used~\cite{bondi1964van} (the radius of hydrogen is changed to 1.1 \AA). Consistent results can be obtained when employing identical settings with Q-Chem. As shown in Figure~\ref{fig:chelpgtime}, in GPU4PySCF, the calculation speed of CHELPG has been greatly improved approximately 10 times versus CPU-based quantum chemistry software~\cite{qchem}. This improvement in computational efficiency is due to the efficient implementation of molecular integrations (calculating the electrostatic potential at grids) on GPU. All the calculations in this subsection are performed using B3LYP functional and def2-SVPD basis with density fitting used.

% The main reason why we implemented CHELPG in GPU4PySCF instead of other charge calculation methods such as MDC (multipole-derived charges) \cite{simmonett2005optimal} is that the charges calculated by CHELPG produce stable atomic charges that align well with experimental data. It is particularly beneficial for research in chemistry, materials science, and biochemistry, where accurate charge distributions are critical for molecular modeling and simulations. For example, as shown in Figure \ref{fig:chelpgcompare}, for the same EC (ethylene carbonate) molecule, CHELPG ensures the transferebility of C atoms between different configurations and the influence of the external environment. However, MDC may have significant changes under the influence of the environment and configuration, resulting in unstable charges (the charge of C atoms ranges from -2.42 to 2.29).
% In summary, the implementation of CHELPG in GPU4PySCF will serve as a powerful tool for obtaining intuitive and quantitative insights into the charge distribution of molecules, playing a crucial role in advancing our understanding of molecular properties and behaviors. All the calculations in this subsection are performed using B3LYP functional and def2-svpd basis with density fitting used.

% \begin{figure}[!htp]
%     \centering
%     \includegraphics[width=0.8\linewidth]{figures/chelpg/comapre.pdf}
%     \caption{Comparison chart of charges on C atoms calculated by CHELPG and MDC for a Li + (EC) 2 cluster. The charges outside the parentheses are calculated by CHELPG, while the charges inside the parentheses are calculated by MDC.
%     \label{fig:chelpgcompare}
%     }
% \end{figure}

%However, the accuracy of the CHELPG method can be contingent upon the level of electronic structure theory applied, as well as the density and distribution of the grid around the molecule. Moreover, the method may face challenges with large molecules where the electron density distribution is highly uneven.

\subsubsection{NMR Shielding Constants}
Nuclear Magnetic Resonance (NMR) Shielding Constants represent a pivotal spectroscopic parameter in NMR spectroscopy, utilized for determining the electronic structure and geometric configuration of molecules~\cite{gunther2013nmr}. During NMR experiments, the absorption signals through nuclear spin transitions under the external magnetic field are detected. Actually, the magnetic field experienced by the nucleus is not the external magnetic field per se, but rather the residual field post electron shielding. The shielding constant elucidates the behavior of the wave function in the vicinity of the nucleus, serving as a coefficient between the electron shielding effect and the external magnetic field. Its ability to provide detailed information about the molecular framework and atomic-level interactions makes it indispensable for chemists. For example, NMR is particularly valuable in organic chemistry for identifying the composition and structure of small organic molecules, aiding in the elucidation of complex organic compounds and reaction mechanisms~\cite{luo2017enantioselective, siuda2019catalyzed}. Employing DFT to calculate NMR shielding constants offers the significant advantage of a level of precision that closely aligns with experimental results with efficiency, yet without the extensive resource demands of experiments~\cite{helgaker2013benchmarking}. This balance makes calculations of NMR shielding constants an invaluable tool for guiding experimental design and interpreting complex NMR spectra, bridging the gap between theoretical insights and practical applications in stereochemistry. In this work, the NMR shielding constants are implemented using the gauge-including atomic orbitals~\cite{london1937theorie} (GIAO) to address the uncertainty of the gauge origin, introduced by the introduction of the uniform external magnetic field. The specific working equations may be referred to in the discussion section on general NMR calculations as presented in Ref.~\cite{yuan2019sublinear}.

Nuclear Magnetic Resonance (NMR) Shielding Constants represent a pivotal spectroscopic parameter in NMR spectroscopy, utilized for determining the electronic structure and geometric configuration of molecules~\cite{gunther2013nmr}. During NMR experiments, the absorption signals through nuclear spin transitions under the external magnetic field are detected. Actually, the magnetic field experienced by the nucleus is not the external magnetic field per se, but rather the residual field post electron shielding. The shielding constant elucidates the behavior of the wave function in the vicinity of the nucleus, serving as a coefficient between the electron shielding effect and the external magnetic field. Its ability to provide detailed information about the molecular framework and atomic-level interactions makes it indispensable for chemists. For example, NMR is particularly valuable in organic chemistry for identifying the composition and structure of small organic molecules, aiding in the elucidation of complex organic compounds and reaction mechanisms~\cite{luo2017enantioselective, siuda2019catalyzed}. Employing DFT to calculate NMR shielding constants offers the significant advantage of a level of precision that closely aligns with experimental results with efficiency, yet without the extensive resource demands of experiments~\cite{helgaker2013benchmarking}. This balance makes calculations of NMR shielding constants an invaluable tool for guiding experimental design and interpreting complex NMR spectra, bridging the gap between theoretical insights and practical applications in stereochemistry. In this work, the NMR shielding constants are implemented using the gauge-including atomic orbitals~\cite{london1937theorie} (GIAO) to address the uncertainty of the gauge origin, introduced by the introduction of the uniform external magnetic field. The specific working equations may be referred to in the discussion section on general NMR calculations as presented in Ref.~\cite{yuan2019sublinear}.

\begin{figure}[!htp]
    \centering
    \includegraphics[width=0.6\linewidth]{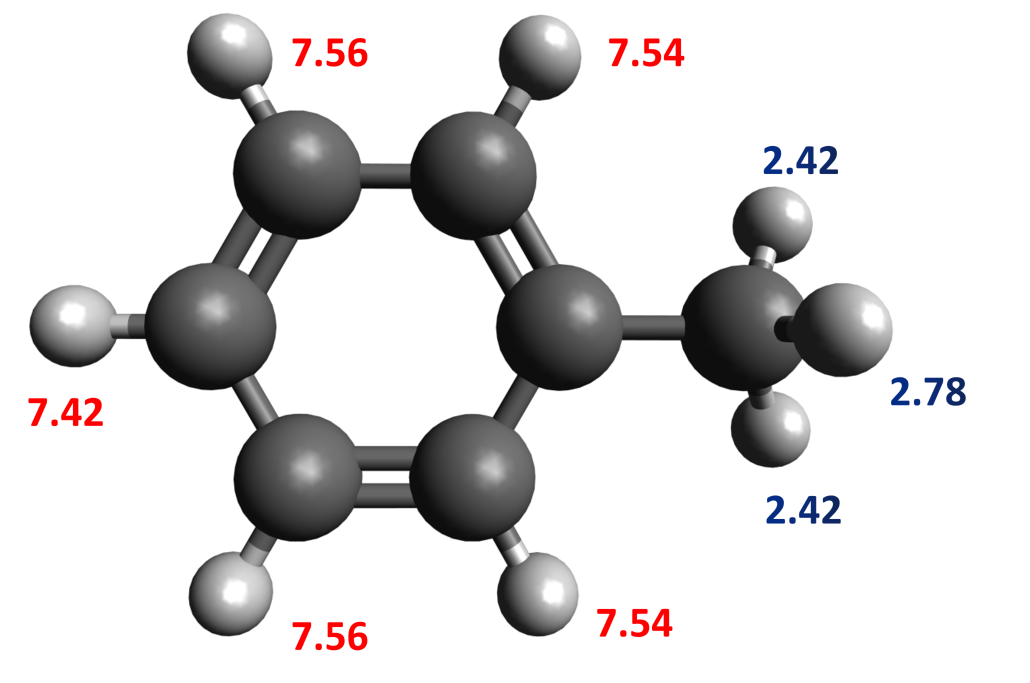}
    \caption{Chemical shifts of hydrogens on toluene (in units of p.p.m.). Hydrogens on the methyl group are indicated in blue text, while hydrogens on the benzene ring are in red. This calculation employs the B3LYP functional, the def2-TZVPP basis set, and the def2-universal-JKFIT auxiliary basis set, and the structure of toluene is taken from PubChem~\cite{kim2023pubchem}.
    \label{fig:nmr}
    }
\end{figure}

An illustrative example of NMR shielding constants of hydrogens on toluene is displayed in Figure \ref{fig:nmr}. When the toluene is oriented perpendicular to the external magnetic field (and for non-perpendicular orientations, it can be decomposed into horizontal and vertical magnetic fields), its delocalized $\pi$ electrons will generate a ring current, which in turn induces a magnetic field. The direction of the induced magnetic field is opposite to that of the external magnetic field on the top and bottom sides of the phenyl ring. However, on the sides of the phenyl ring (where the hydrogen atoms are positioned on the sides of the ring), the directions are the same, i.e., the induced magnetic field enhances the effect of the external magnetic field, deshielding the hydrogen nuclei and causing the chemical shift to move to lower field values; whereas the hydrogen atoms on the methyl group are in a stronger shielding environment, leading to the chemical shift to move to higher field values. Due to the geometrical configuration, the three hydrogen atoms on the methyl group exhibit differing chemical shifts, although theoretically, they should be identical.

% In toluene, the hydrogen atoms on the benzene ring experience a descreening effect due to the local magnetic field effect produced by the $\pi$ electron cloud, resulting in their shielding constants being relatively low\cite{gunther2013nmr}; whereas the hydrogen atoms on the methyl group are in a stronger shielding environment, leading to higher shielding constants. Consequently as shown in Figure \ref{fig:nmr}, the hydrogen atoms on the benzene ring exhibit a higher chemical shift relative to those on the methyl group, reflecting the direct outcome of the descreening effect.
% \wxj{add description for three different H's}

% The theoretical computation of shielding constants can be performed through non-relativistic and relativistic approaches. While the non-relativistic methodology is relatively mature, it often yields inaccurate results for molecules containing heavy elements. In essence, NMR shielding constants mirror the state of electrons surrounding the nucleus, constituting an essential conduit between theoretical calculations and experimental observations. Through the calculation of shielding constants, a deeper understanding of molecular electronic structures is attainable, facilitating the prediction of chemical shifts.

\subsection{Solvation Free Energy}
\label{sec:solvation_free_energy}
Implicit solvent models are computational tools designed to efficiently simulate the effect of a solvent on molecular systems, emphasizing the solvent's overall impact rather than detailing every interaction between solvent and solute molecules. In the realm of quantum chemistry, electrostatic interactions between solvent and solute molecules are captured using the Polarizable Continuum Model (PCM), which offers a nuanced understanding of molecular properties in solution. In our enhancement of GPU4PySCF, users can access four variants of PCM models: C-PCM, COSMO, IEF-PCM, and SS(V)PE. For accurate prediction of solvation energies, accounting for non-electrostatic interactions such as cavitation, Pauli repulsion, dispersion, and hydrogen bonding is essential. These contributions are encompassed within more advanced models like SMD~\cite{SMD}, CMIRS~\cite{cmirs}, and COSMO-RS~\cite{cosmo-rs} models, with the SMD model being particularly prevalent in quantum chemistry calculations and available from GPU4PySCF.

The implementation of the SMD model varies slightly across different quantum chemistry software platforms. And the algorithm has been updated with minor modifications over time. Compared with other implementations in Q-Chem and G03, we point out several {\it possible} differences in our implementation:
\begin{itemize}
\item The cavity surface of PCM models is smoothed using SWIG methods~\cite{swig-pcm}, enhancing the stability of geometry optimization and molecular dynamics simulations.
\item Atomic SASA, molecular SASA, and CDS contributions are calculated using the Fortran code from NWChem \cite{nwchem}. Those calculations are also re-implemented in Python, where atomic SASA and molecular SASA are calculated numerically using Lebedev quadrature, diverging from the analytical formulation in \cite{sasa}. The difference is usually negligible for calculating the CDS contribution.
\item Modified Coulomb radii are employed with SMD18~\cite{SMD18}, with newly fitted parameters that improve the accuracy of halogen bonding interactions.
\end{itemize}

We employ the Minnesota Solvation Database – version 2012~\cite{marenich2020minnesota} for the benchmarks. The database has been updated to version 2020 since the publication of SMD. The corresponding solvent descriptor database has been updated twice. We should take the changes into consideration for the following results. For neutral solutes, the error in solvation free energies consistently remains below 1 kcal/mol across various protocols (Table \ref{tab:smd}), with the errors decreasing further when employing larger basis sets or advancing up Jacob's Ladder. These errors are comparable to those from corresponding protocols in Q-Chem and Gaussian 03. For achieving the best accuracy, a protocol employing a larger basis set and a higher-level Rung exchange-correlation functional is recommended. For ions, the overall error in solvation free energies is around 4 kcal/mol, with cases involving acetonitrile and dimethyl sulfoxide contributing significantly to this discrepancy. Notably, SMD calculations using Hartree-Fock generally outperform those utilizing DFT methods such as B3LYP and M06-2X, a conclusion supported by other studies~\cite{pcm_review,smd_benchmark0,SMD}.

\begin{table}[!htp]
    \centering
    \begin{tabular}{c|cc|cccc}
        \hline
        & \multicolumn{2}{|c|}{Neutrals} & \multicolumn{4}{c}{Ions} \\
         Method & Aqueous  & NAQ & Acetonitrile & DMSO & Methanol & Water \\
         \hline
         GPU4PySCF/M06-2X/TZ & 0.72 & 0.68 & 6.0 & 4.2 & 2.4 & 5.5\\
         GPU4PySCF/M06-2X/6-31G* & 0.74 & 0.65 & 5.9 & 4.6 & 2.1 & 4.8\\
         GPU4PySCF/B3LYP/TZ & 0.82  & 0.68 & 5.7 & 3.8 & 3.1 & 6.2\\
         GPU4PySCF/B3LYP/6-31G*  & 0.94  & 0.70 & 5.8 & 4.3 & 2.5 & 5.4\\
         GPU4PySCF/HF/6-31G* & 0.91 & 0.74 & 6.2 & 5.4 & 3.1 & 3.8\\
         Q-Chem/HF/6-31G* & 0.90 & 0.70 & \multicolumn{3}{c}{5.4/4.1} & 2.9/3.9 \\
         G03/M05-2X/cc-pVTZ & 0.68 & 0.67 & 5.9 & 4.4 & 2.3 & 4.6\\
         G03/B3LYP/6-31G* & 0.80 & 0.67 & 5.7 & 4.3 & 2.9 & 4.9\\
         G03/HF/6-31G* & 0.90 & 0.73 & 5.9 & 5.2 & 2.7 & 3.4 \\
         \hline
    \end{tabular}
    \caption{Mean unsigned error (MUE) in solvation free energies (kcal/mol) with different protocols. The detailed protocols for G03 are described in the original SMD paper~\cite{SMD}, Table 13. The data for Q-Chem is take from~\cite{pcm_review}.   In all calculations, the bulk electrostatic contribution is calculated with IEF-PCM. Density fitting is applied for all GPU4PySCF calculations. 2346 solvation free energies for neutrals and 332 solvation free energies for ions are used for G03~\cite{SMD} 2368 solvation free energies for neutrals and 362 solvation free energies for ions are used for PySCF results.  The uncertainty of the reference data are $\pm$ 0.2 kcal/mol for neutral solutes and $\pm$ 3 kcal/mol for ions. `TZ' is short for def2-TZVPP. `NAQ' is short for nonaqueous. }
    \label{tab:smd}
\end{table}

\subsection{Chemical Reactions}
\subsubsection{Transition State Search}
\label{sec:ts}
The activation energy, the energy of the transition state relative to the reactants, is a critical parameter for determining the chemical reaction rate. A higher activation energy means the fewer molecules have enough energy to react at a given temperature, leading to a slower reaction. Knowing the activation energy helps in designing and optimizing chemical processes. Transition state searches in quantum chemistry are complex and require careful planning and execution. Computational tools play an important role for finding the optimized geometries of the reactants and products, optimizing the transition state, and verifying the transition state. Those computations need a lot of computational resources for calculating SCF, gradient, and Hessian. An efficient quantum chemistry tool can accelerate the finding of the reaction pathway.
Successfully finding transition states requires not just computational resources but also a deep understanding of the chemistry involved and experience with the computational methods and software. There are fully or partially automated tools~\cite{ts_xtb,auto_de,auto_ts} for generating reaction profiles. This work does not provide the automated approaches. But one can build GPU4PySCF in the existing automated workflows.

In this work, we try to reproduce the result in MOBH35 dataset~\cite{MOBH35}. For simplicity, the reactions involving bimolecular reactants or products are removed~\cite{ts_xtb}. This subset of MOBH35 includes various transition metals Sc, Ti, Mn, Fe in the first row, Nb, Mo, Ru, Rh, Pd in the second row, and Ta, W, Re, Os, Ir, Pt in the third row. The transition metals are generally challenging for computation because they not only requires high-angular momentum GTO integrals, but also need ECP in the large basis set. Ref.~\cite{ts_xtb} tried to modify the geometries of some reactions based on the chemical intuition. However, we are not able to verify the transition states in the modified geometries. We still use the original geometries in~\cite{MOBH35} as the initial guess for re-optimizing transition state. Both geometry optimizations and transition state optimizations are performed with geomeTRIC~\cite{geometric}.

\begin{figure}
    \includegraphics[width=1.0\linewidth]{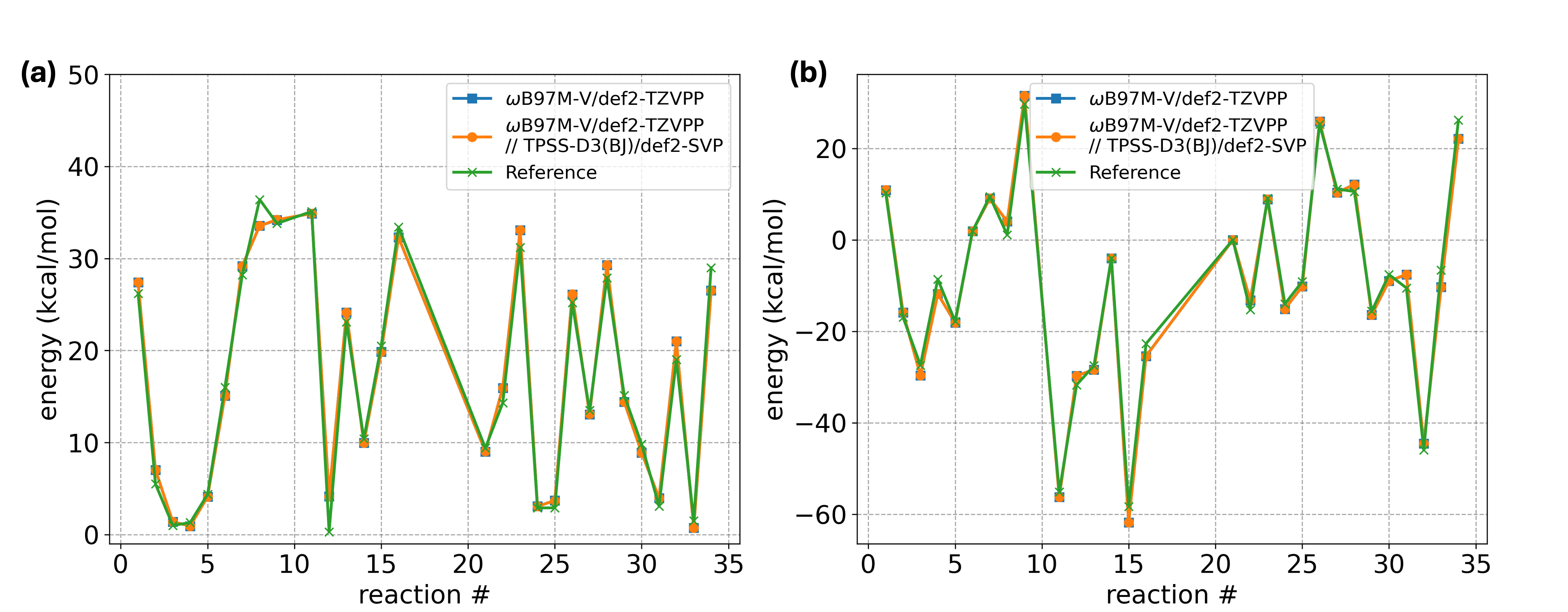}
    \caption{Activation energy (left) and reaction energy (right). $\omega$B97M-V/def2-TZVPP: direct single-point energy calculations with the geometries in Ref.~\cite{MOBH35} using $\omega$B97M-V functionals and def2-TZVPP basis set. $\omega$B97M-V/def2-TZVPP//TPSS-D3(BJ)/def2-SVP: re-optimize geometries using the same protocols as~\cite{MOBH35} TPSS-D3(BJ)/def2-SVP and calculate the single-point energy with using $\omega$B97M-V functionals and def2-TZVPP basis set. All the calculations employ (200,1202) grids for transition metals, (99,590) grids for other elements, and density fitting.}
    \label{fig:ts}
\end{figure}

We benchmark the results with $\omega$B97M-V, which is one of the best performer in MOBH35 database~\cite{MOBH35}. In the dataset, the reference energies are calculated with $\mathrm{CCSD(T)/CBS_{w1}}$ + $\mathrm{\Delta(T)/TZVPP}$. The barrier heights and reaction energies are presented in Fig.~\ref{fig:ts}. The average absolute deviation (MAD) of the reaction energy we calculated in the subset of MOBH35 is 1.75 kcal/mol, which is similar to MAD (1.7 kcal/mol) of the entire database reported in \cite{MOBH35}. The re-optimized transition states are almost identical to the geometries reported in \cite{MOBH35}. The difference of the reaction energy between the original geometry and re-optimized geometries is 0.01 kcal/mol in MAD. That verifies the robustness of transition state optimization scheme.

\subsubsection{Electrochemical Stability Windows of Electrolytes}
\label{sec:redox}
To guarantee thermodynamic stability of a lithium-ion battery, it's critical that the electrochemical potential of the electrodes falls within the electrolyte's electrochemical stability window, which is the range between the oxidation and reduction potential of the electrolyte~\cite{marchiori2020understanding,delp2016importance,hall2018exploring}. If an anode's electrochemical potential exceeds the reduction potential of the electrolyte, it can lead to the reduction of the electrolyte; similarly, if a cathode's electrochemical potential is below the oxidation potential of the electrolyte, it could oxidize the electrolyte, despite that the formation of a passivation layer can prevent continuous electron transfer and mitigate these effects. A broader electrochemical window signifies a larger operational voltage range, which directly translates to higher energy density and potentially higher power output of the device~\cite{goodenough2013li,goodenough2012rechargeable}. Thus, in the study of electrolyte in lithium-ion batteries, electrolyte's electrochemical window directly influences the selection of solvent molecules in electrolyte for high-performance energy storage solutions.

A straightforward approach to determining the electrochemical stability window of electrolytes involves calculating the gap between HOMO and LUMO~\cite{goodenough2013li,goodenough2012rechargeable}, but this method has its shortcomings~\cite{peljo2018electrochemical}. Thus, our analysis adopts a more accurate thermodynamic approach to define the electrochemical stability of the electrolyte, which relies on calculating the redox potentials rather than molecular orbital gaps~\cite{hall2018exploring,borodin2013oxidative,delp2016importance,hou2019influence,borodin2015towards,peljo2018electrochemical}.
The determination of oxidation and reduction potentials can be rigorously approached through thermodynamic cycles~\cite{borodin2013oxidative,peljo2018electrochemical}. Besides, for the evaluation of reduction potentials in particular, the introduction of $\mathrm{Li^+}$ ions plays a pivotal role for considering the polarization effects of $\mathrm{Li^+}$ on solvent molecules, which significantly influence the electrochemical behavior of the system, thereby affecting the solvent's reduction potential~\cite{borodin2015towards,hall2018exploring}. Thus, in this work, we consider the following oxidation and reduction reactions for some specific solvent $ \mathrm{S} $
\begin{align}
    \mathrm{S}\rightarrow\mathrm{S}^++\mathrm{e}^{-}, \\
    \mathrm{S}+\mathrm{Li}^++\mathrm{e}^{-}\rightarrow\mathrm{Li}-\mathrm{S}.
\end{align}
The oxidation and reaction potential potential of those reactions can be calculated as
\begin{align}
    E_{\mathrm{oxidation}} = &(G_{\mathrm{gas}}(\mathrm{oxidized})-G_{\mathrm{gas}}(\mathrm{initial})+\Delta G_{\mathrm{solv}}(\mathrm{oxidized})-\Delta G_{\mathrm{solv}}(\mathrm{initial}))/F-1.44, \label{equ:oxi} \\
    E_{\mathrm{reduction}} = & -(G_{\mathrm{gas}}(\mathrm{reduced})-G_{\mathrm{gas}}(\mathrm{initial})+\Delta G_{\mathrm{solv}}(\mathrm{reduced})-\Delta G_{\mathrm{solv}}(\mathrm{initial}))/F-1.44, \label{equ:red}
\end{align}
where $G_{\mathrm{gas}}$ is the free energy in gas of different species, such as initial, oxidized and reduced species. The solvation energy in Eq. (\ref{equ:oxi}) and Eq. (\ref{equ:red}) of some specific solvent $ \mathrm{S} $ reads
\begin{align}
    \Delta G_{\mathrm{solv}}(\mathrm{S}) =E_{\mathrm{solv
}}(\mathrm{S})-E_{\mathrm{gas
}}(\mathrm{S}),
\end{align}
approximated by the energy difference between solvation and gas phase. Subtraction of 1.44 V converses from the absolute electrochemical scale to the $\mathrm{Li/Li^+}$ potential scale, and the value 1.44 V is referred to Ref.~\cite{hall2018exploring} for comparisons.

\begin{figure}[!htp]
    \centering
    \includegraphics[width=0.8\linewidth]{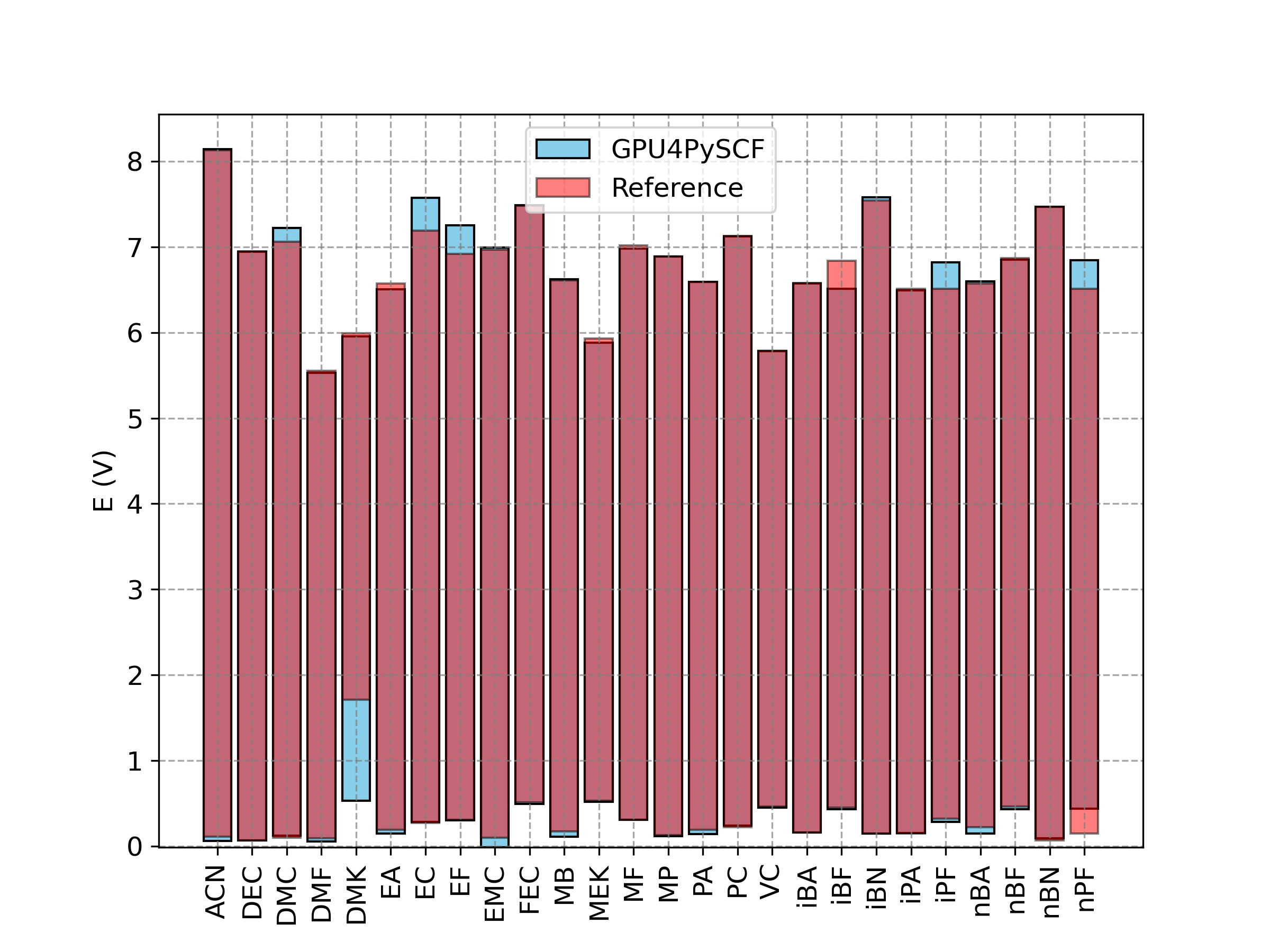}
    \caption{Electrochemical stability windows calculated by GPU4PySCF (blue bars) and from Ref.~\cite{hall2018exploring} (red bars). It should be noted that the largest discrepancy lies in the reduction potential of the DMK molecule. Using the same structure and calculation method as Ref.~\cite{hall2018exploring}, we get 0.55 V, which differs from the reported results. Computational details are the same as in section \ref{subsec:battery_cluster}.
    \label{fig:ecwindow}
    }
\end{figure}

In our study, we calculated the redox potentials of some molecules reported in Ref. \cite{hall2018exploring}, and compared our results with those previously reported in the same reference. It should be noted that our calculations were performed using a thermodynamic cycle, with computational parameters that slightly differ from those used in Ref. \cite{hall2018exploring}. As shown in Figure \ref{fig:ecwindow}, despite differences in the computational approach and parameters from those described in the original literature, the results obtained are remarkably similar. Consequently, we posit that the package possesses the capability to calculate redox potentials, thereby enabling the prediction of electrochemical windows. Considering the computational efficiency, oxidation and reduction potential calculations involves the optimization of geometric structures in vacuum and solvent, as well as the calculation of Hessian in vacuum. The computational acceleration using GPU4PySCF with respect to Q-Chem for these processes has been discussed in previous sections. Consequently, the acceleration ratio for the calculation of oxidation-reduction potential is determined by the lowest acceleration ratio among these individual parts.
% \wxj{Add some details of computational cost?}

\subsubsection{Fukui Function, Condensed Fukui Function and ESP}
\label{sec:fukui}
In chemical reaction research, determining the most reactive site on a compound is crucial. Typically, organic chemists make judgments based on knowledge and experience, but quantum chemistry calculations can also be helpful. The widely-used Fukui function~\cite{lu2012multiwfn,parr1984density} predicts reactive sites and comprises three types: \emph{$f^{+}$}, \emph{$f^{-}$}, and \emph{$f^{0}$}, for nucleophilic, electrophilic, and radical attacks, respectively. In most applications, the Fukui function is calculated by the finite difference formula~\cite{flores2008efficient}. It requires calculations for neutral (N electrons), cationic (N-1 electrons), and anionic (N+1 electrons) states and analyzing the electron density difference. Regions with larger Fukui function values imply greater reactivity at this site.

The condensed Fukui function uses atomic charges to evaluate the reactivity of individual atoms in a molecule more intuitively~\cite{yang1986use}. By calculating atomic charges for N, N-1, and N+1 electron states, differences between states can be obtained for each atom, enabling reactivity comparisons.

For example, consider an electrophilic aromatic substitution using Anisole to examine the orientational effect of the directing group. After optimizing the initial structure for the neutral state, single point calculations are performed for N, N-1, and N+1 electron states at the B3LYP/def2-TZVPP level. The Fukui function is obtained from electron density differences in different states and saved to *.cub files, which can be visualized using the open-source tool VMD~\cite{humphrey1996vmd}. Finally, the condensed Fukui function is calculated using CHELPG atomic charges~\cite{breneman1990determining}.
\begin{figure}[!htp]
	\centering
	\includegraphics[width=0.5\linewidth]{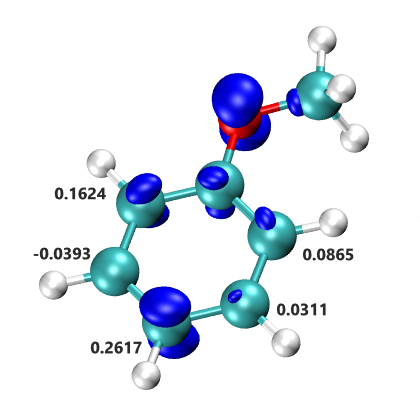}
	\caption{
		\label{fig:fukuifunc}
		The \emph{$f^{-}$} Fukui function isosurface (isolevel 0.008) of Anisole is shown in blue, and the condensed \emph{$f^{-}$} values are marked near the corresponding carbon atoms.
	}
\end{figure}

As the study concentrates on electrophilic substitution sites, only the \emph{$f^{-}$} function is relevant. As shown in Figure~\ref{fig:fukuifunc}, the \emph{$f^{-}$} isosurface (isolevel 0.008) occurs mainly on the \emph{ortho}- and \emph{para}-position carbon atoms, and less in the \emph{meta}-position. The condensed Fukui function values also suggest that \emph{ortho} and \emph{para} carbon atoms with higher \emph{$f^{-}$} values are more susceptible to electrophilic substitution reactions. This result is consistent with organic chemistry knowledge that the -OCH$_{3}$ group is an \emph{ortho}, \emph{para} director.

\begin{figure}[!htp]
	\centering
	\includegraphics[width=0.5\linewidth]{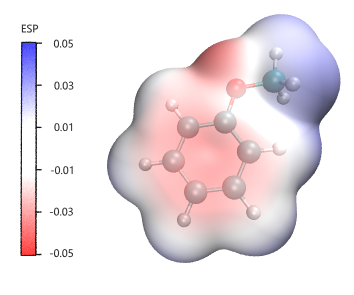}
	\caption{
		\label{fig:espfunc}
		ESP color map for Anisole on the electron density isosurface (rho=0.001).
	}
\end{figure}

The electrostatic potential (ESP) is a commonly used real-space function that provides important information about the interaction sites of compounds. It helps understand the anisotropic distribution of charges on the molecular surface, thereby assisting in inferring possible reaction sites. The wave function obtained from the single-point calculation of compounds using GPU4PySCF can be further exported as the ESP's *.cub file, which can also be visualized using VMD. The color map of the electrostatic potential for Anisole on the electron density isosurface (rho=0.001) is displayed in Figure~\ref{fig:espfunc}.

\subsection{Incorporating with Neural Networks}
 Recently, the application of machine learning has been extended to the quantum chemistry community \cite{ferminet,dm21}, demonstrating substantial potential to address quantum mechanic problems. The traditional quantum chemistry techniques are still important ingredients of the recipe. Currently, most researchers employ the PySCF routines for accessing the quantum chemistry calculations. This work further offers the same functionalities with GPU acceleration, which is crucial as most deep learning models are running on GPU.

\begin{figure}[!htp]
    \centering
    \includegraphics[scale=0.6]{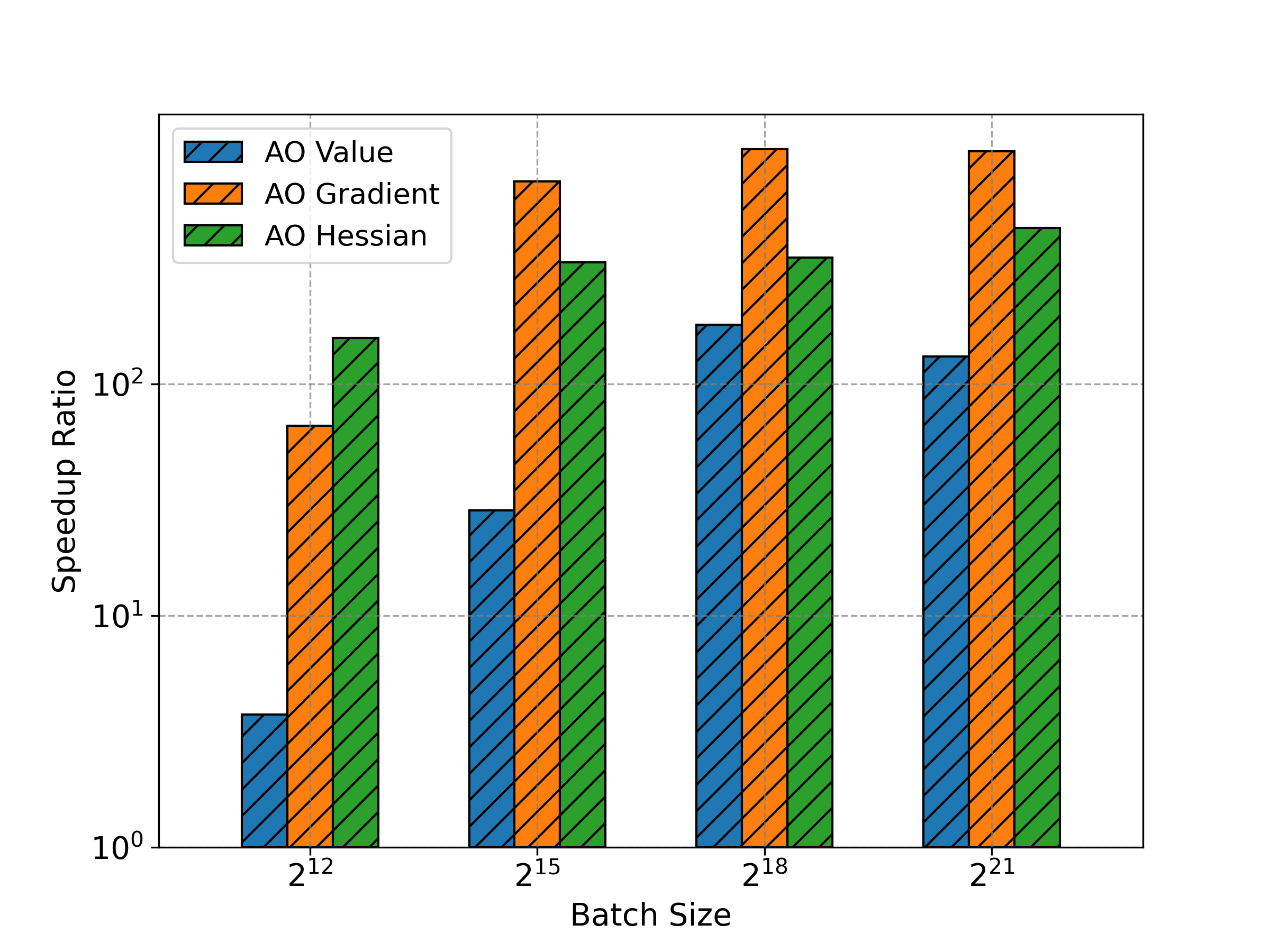}
    \caption{Speedup ratio of A100-80G card over 32-core CPUs in atomic orbital evaluations. Benzene with cc-pVDZ basis is used for benchmark. AO values, gradients and Hessian matrices are chosen for comparison.}
    \label{fig:ao_speedup}
\end{figure}

We take Fermionic neural networks \cite{deepwf, paulinet, ferminet} as an example. The main philosophy of this research realm, namely neural network-based Quantum Monte Carlo, is to learn highly accurate ground state wavefunctions using a variational Monte Carlo approach. For this approach, atomic orbital (AO) evaluation is an important component that encapsulates traditional chemistry knowledge. Specifically, utilizing the Hartree-Fock AO to `pretrain' neural networks is a common strategy in this research area, as introduced in \cite{ferminet}. This process guides the networks to an acceptable initial state, after which they evolve based on variational principles. However, AO evaluation, invoked at every `pretraining' step, is typically executed on CPUs, which quickly becomes the computational bottleneck of the whole process. Another factor contributing to the inefficiency of this implementation is substantial data transfer between CPU memory and GPU memory, especially for large systems. With GPU4PySCF developed, AO evaluation can be efficiently performed on GPUs and achieves a hundredfold speed improvement as shown in Fig.~\ref{fig:ao_speedup}.

Furthermore, GPU4PySCF can be effortlessly integrated with existing neural network frameworks within quantum chemistry and manifest significant acceleration. For instance, by integrating LapNet~\cite{lapnet} with GPU4PySCF, we achieved approximately three times faster pretraining on Benzene molecule. Beyond pretraining, GPU4PySCF also enables researchers to design and implement neural network wavefunction ansatzes that incorporate traditional chemistry knowledge, like AO, to better characterize the desired quantum state.
% \wxj{the statement `we believe' is not scientific enough. Can we add some supportive details, instead of claims?}

\section{Compatibility and Functionalities}
\label{sec:community}
%PySCF has been a constantly growing open-source community. It has been integrated into various open-source packages as the interface to quantum chemistry. For example,
%\begin{itemize}
%    \item {\bf Quantum Computing}. For instance, Google's quantumlib, IBM's Qiskit, NVIDIA's CUDA Quantum, Tencent's TenCirChem.
%    \item {\bf Deep Learning}. For instance, DeepMind's FermiNet~\cite{ferminet} and DM21, DeepModeling's DeePKS-kit~\cite{deepks}, ByteDance's DeepSolid and LapNet.
%    \item {\bf Quantum Chemistry Workflows}. For instance, Microsoft's AiiDA.
%\end{itemize}
As an open-source Python library, PySCF has been widely adopted in quantum computing, deep learning, and various quantum chemistry workflows, most of which extensively utilize GPU acceleration. In this section, we introduce improvements to the interfaces and outline the functionalities that have been accelerated. With minimal code changes, these tasks can be sped up using GPU4PySCF v1.0.

\subsection{Improved Compatibility with PySCF}
\lstdefinestyle{mystyle}
{
    language = Python,
    %basicstyle = {\ttfamily \color{main-color}},
    %backgroundcolor = {\color{back-color}},
    %stringstyle = {\color{string-color}},
    %keywordstyle = {\color{key-color}},
    keywordstyle = [2]{\color{red}},
    otherkeywords = {to_gpu},
    morekeywords = [2]{to_gpu},
}
\label{sec:compatibility}
The compatibility between PySCF and GPU4PySCF has been enhanced by the newly designed API in this work. Previously, all GPU4PySCF classes inherited from PySCF classes, with methods patched to use GPU functions. As more functionalities were added, the multiple inheritance design leads to many ambiguous situations when the features are defined in both PySCF and GPU4PySCF. The ambiguity increases the complexity of development dramatically. Starting from v1.0, GPU4PySCF drops the patching mechanism and explicitly redefines the same classes as in PySCF. GPU4PySCF classes and PySCF classes are constructed in parallel. A PySCF object can be converted into a GPU4PySCF object with calling
\lstinline[language=Python]!to_gpu! function when the functionality is implemented on GPU. The  \lstinline[language=Python]!kernel! execution will be accelerated with GPU. Conversely, a GPU4PySCF object can be converted back into a PySCF object using the \lstinline[language=Python]!to_cpu! function, allowing PySCF-only features to be applied. Here is an example

\begin{lstlisting}[language=Python,
    style=mystyle,
    caption=Example of PySCF with GPU support. The code changes from CPU to GPU are colored in red.]
import pyscf
from pyscf import lib
from pyscf.dft import rks

atom = '''
O       0.0000000000    -0.0000000000     0.1174000000
H      -0.7570000000    -0.0000000000    -0.4696000000
H       0.7570000000     0.0000000000    -0.4696000000
'''

mol = pyscf.M(atom=atom, basis='sto3g')
mol.verbose = 4
mf = rks.RKS(mol, xc='b3lyp').density_fit()

# Move PySCF object to GPU
mf_GPU = mf.to_gpu()

# Compute Energy on GPU
e_dft = mf_GPU.kernel()

# Compute Gradient
g = mf_GPU.nuc_grad_method()
g_dft = g.kernel()

# Compute Hessian on GPU
h = mf_GPU.Hessian()
h.auxbasis_response = 2
h_dft = h.kernel()
\end{lstlisting}

%\subsection{High consistency with PySCF on CPU}
With the new design, GPU4PySCF v1.0 has been meticulously engineered to maintain a high degree of compatibility with PySCF, ensuring that any settings related to accuracy in PySCF are seamlessly applicable in GPU4PySCF. For example, consider DFT calculations.

We have conducted benchmarks on SCF energy, gradient, and Hessian calculations for the water molecule using well-established exchange-correlation functionals across Rungs 2-4 of Jacob's Ladder~\cite{jacobs_ladder}. For all examined exchange-correlation functionals, the discrepancies in energy relative to PySCF calculations are less than $10^{-11}$ Ha, falling below the SCF convergence threshold of $10^{-10}$. Moreover, the 2-norm differences for gradients and Hessians are below $10^{-7}$ $\text{Ha/Bohr}$ and $10^{-6}$ $\text{Ha/Bohr}^2$, respectively, demonstrating exceptional accuracy and consistency with PySCF. The large differences in range-separated functionals are due to solving ill-conditioned linear system in density fitting.
%GPU4PySCF is designed to be highly consistent with PySCF. Any PySCF settings involving accuracy are reused in GPU4PySCF. The consistency can be easily achieved with the compatible APIs, which will be discussed in Sec. \ref{sec:compatibility}. And the customized config will also be applied to GPU4PySCF automatically. We benchmark SCF energy, gradient and Hessian calculations of water molecule with well-known exchange-correlation functionals on Rung 2-4 of Jacob's ladder~\cite{jacobs_ladder}. For all XC functionals, the energy differences from PySCF calculations are smaller than $10^{-11}$ Ha, which is less than SCF convergence tolerance $10^{-10}$. The 2-norm of gradient difference and Hessian difference are smaller than $10^{-7}$ $\text{Ha/Bohr}$ and $10^{-6}$ $\text{Ha/Bohr}^2$ respectively.

\begin{figure}[!htp]
    \centering
     \includegraphics[scale=0.5]{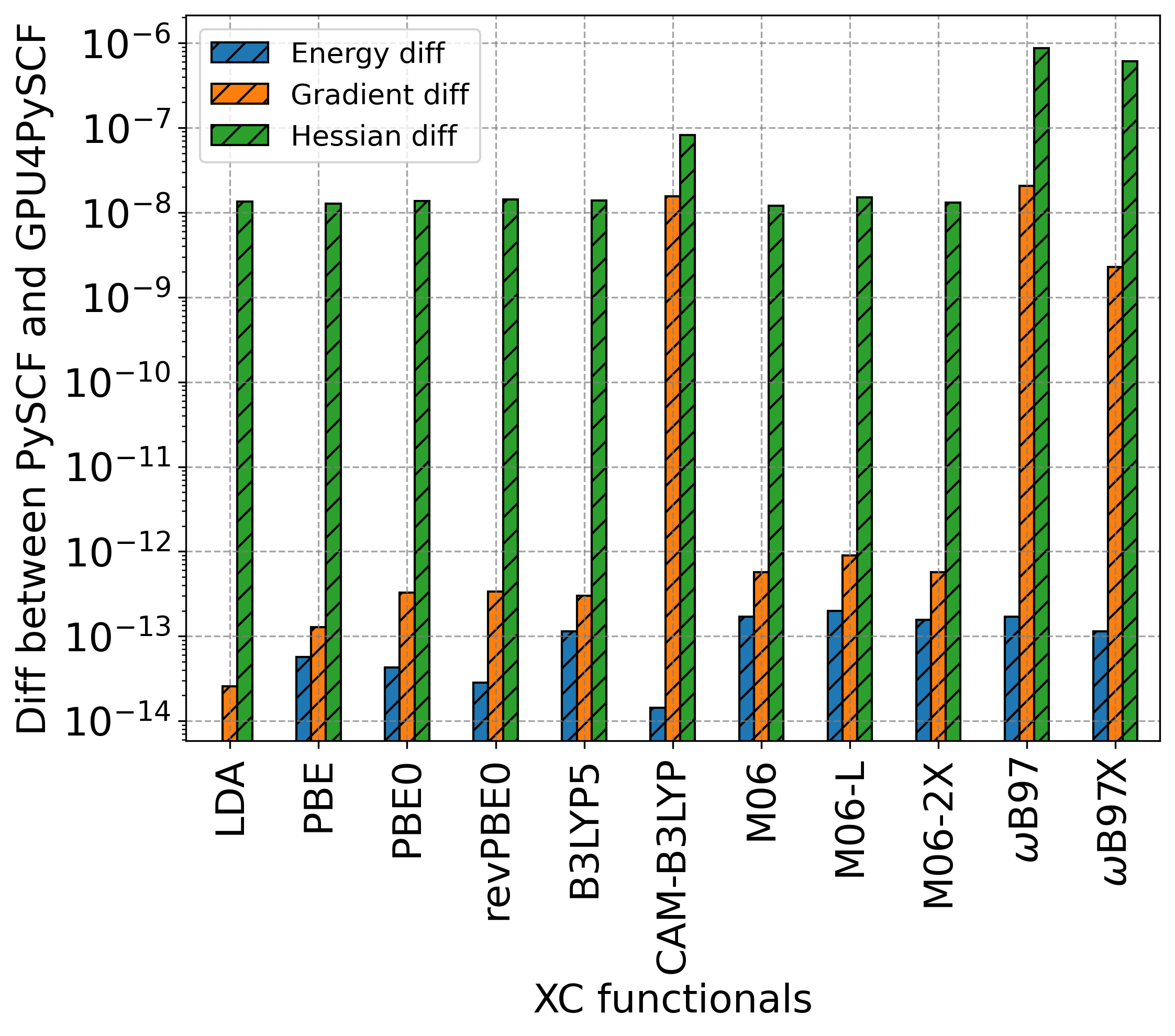}
     \caption{2-norm of the differences between PySCF and calculations with different DFT functionals for $\mathrm{H_2O}$ molecule, grids level = 5, def2-QZVPP atomic basis, def2-universal-JKFIT auxiliary basis, convergence tolerance = $10^{-12}$. Different units (Ha for energy, Ha/Bohr for gradient, and $\text{Ha/Bohr}^2$ for Hessian) are used in y-axis.}
\end{figure}

%At the developer level, GPU4PySCF follows the same Python function and class names as PySCF, such that modules in PySCF can be reused. However, the low-level implementations are significantly different, such as GTO integrals, DFT integrals, and JK kernels.

\subsection{Accelerated Functionalities}
\label{sec:functionalities}
Practical DFT calculations not only involve the contributions of foundational DFT but also advanced features such as dispersion correction, solvent effect, ECP, and etc. Those contributions are essential to accurately model the electronic structure. This study adheres to a principle of utilizing existing open-source implementations for these advanced features wherever possible. In scenarios where these implementations are either non-existent or inadequate, we undertake the redevelopment of these features, emphasizing the enhancement of computational efficiency through GPU acceleration, particularly for processes that exhibit slower performance than SCF iterations.

Dispersion corrections are usually much cheaper than DFT self-consistent iterations. And these contributions are calculated with dftd3/simple-dftd3 and dftd4/dftd4 packages without GPU acceleration. Although DFT-D3 has been implemented in Torch and Jax, these implementations will introduce the dependencies of heavy-weight machine learning framework. For robustness, we rebuild simple-dftd3 and dftd4, and disable OpenMP. The same interfaces are used in both PySCF and GPU4PySCF.  Nonlocal corrections are carefully optimized at CUDA level due to its expensive computational cost.

PCM/SMD solvent models require solving Poisson-Boltzmann equation in each SCF iteration. The computational complexity of solving Poisson-Boltzmann equation is cubic-scaling with respect to the number of grids on molecular surface. Although it is cheaper than the quartic-scaling hybrid DFTs, it would be the bottleneck without GPU acceleration. To address this, we have re-engineered the PCM/SMD solvent models for GPU execution using Python, recognizing the potential need for further optimizations to minimize the memory requirements. To maintain the compatibility with PySCF we also introduced PCM/SMD solvent models in PySCF v2.4 and added nuclear Hessian in PySCF v2.6.

ECP contributions are calculated only once in SCF. We reuse PySCF routines on CPU to calculate the ECP contributions in SCF, gradient, and Hessian. Other contributions of DFT are still accelerated with GPU for a molecular system with ECP. The computational cost is acceptable if the system only has a few heavy atoms. We leave the GPU acceleration of ECP as future work.

The functionalities are summarized in the Table~\ref{tab:func}.
\begin{table}[!htp]
\centering
\begin{tabular}{c|ccc}
\hline
Method & SCF & Gradient & Hessian \\
\hline
direct SCF \cite{gpu4pyscf-caltech} & O & O & CPU\\
density fitting & O & O & O \\
LDA & O & O & O \\
GGA & O & O & O \\
mGGA &  O & O & O \\
hybrid & O & O & O \\
unrestricted & O & O & O\\
PCM solvent & GPU & GPU & FD \\
SMD solvent & GPU & GPU & FD \\
dispersion correction & CPU* & CPU* & FD \\
nonlocal correlation & O & O & NA\\
ECP & CPU & CPU & CPU \\
MP2 & GPU & CPU & CPU \\
CCSD & GPU & CPU & NA \\
\hline
\hline
\end{tabular}
    \caption{A summary of GPU4PySCF functionalities and level of optimizations. `O': carefully optimized for GPU. `CPU': only cpu implementation. `GPU': drop-in replacement or na\"ive implementation. `FD': use finite-difference gradient to approximate the exact Hessian matrix. 'NA': not available. `CPU*': DFTD3~\cite{dftd3}/DFTD4~\cite{dftd4} on CPU.}
    \label{tab:func}
\end{table}

\subsection{Improved Interfaces to External Libraries}
%It also has been incorporated with other open-source packages, such as geomeTRIC~\cite{geometric} and Berny for geometry optimization, dftd3\cite{dftd3} and dftd4\cite{dftd4} for dispersion corrections, LibXC\cite{libxc} for newly developed exchange-correlation functionals, basis set exchange~\cite{bse} for newly developed basis set.
%The development of PySCF/GPU4PySCF benefits from the contributions from the open-source community.  GPU4PySCF, acting as an extension of PySCF, can be seamlessly incorporated with the open-source packages compatible with PySCF. We highlight the roles of the following fundamental packages working with PySCF/GPU4PySCF.
One of the advantages that we have benefited from by working with PySCF and GPU4PySCF is that we can benefit from the contributions from the open-source community. In particular, GPU4PySCF, acting as an extension of PySCF, can be seamlessly incorporated with the open-source packages compatible with PySCF. In addition to the many libraries that PySCF depends on BSE~\cite{bse} (basis set exchange for newly developed basis set), GeomeTRIC~\cite{geometric} (geometric optimization and transition state search), DFTD3~\cite{dftd3}/DFTD4~\cite{dftd4} (dispersion corrections) and Libcint~\cite{libcint} (fundamental GTO integrals), our work in GPU4PySCF has benefited extensively from the CUDA ecosystem:
\begin{itemize}
    \item CuPy~\cite{cupy}, a drop-in replacement for NumPy \& SciPy for GPU. 
    \item CUDA LibXC. CUDA support is still experimental in the latest version of libXC (v6.2). In this work, the exchange-correlation potentials and their derivatives are evaluated on the GPU using a customized Python API.
    \item cuTENSOR. cuTENSOR is used as the default tensor contraction engine. It takes the advantage of tensor cores, and provides convenient tensor operations on NVIDIA GPUs. Other tensor contraction packages, such as opt\_einsum~\cite{Smith2018} and cuquantum~\cite{cuquantum} are also supported.
\end{itemize}

% Add
% - We used external librairies including .... 
% - particularly, additional effort (reasons) 
% - Developing CUDA python wrapper for cuda libxc backend

\section{Limitations and Future Development}
\label{sec:limitations}
Quantum chemistry is a field characterized by its complex algorithms and significant computational demands. Accelerated computing technologies, such as GPUs, offer promising ways to tackle these challenges, yet their adoption in quantum chemistry has been slower than in other areas. With our many contributions to the GPU4PySCF package, we can now deliver significant speedup for various tasks. As demonstrated through our applications, GPU4PySCF can now be applied in many different practical settings, bringing the full power of GPUs to bear on a large part of quantum chemistry modeling. We can thus recommend GPU4PySCF as a powerful module for the PySCF ecosystem in industrial applications.

At the same time, our implementations are certainly not optimal and there are many current limitations. As dedicated stakeholders in this project, we outline our vision on some of the important future developments to further enhance the capabilities of this package.

{\bf Wavefunction-based Methods}. Beyond traditional mean-field methods, sophisticated wavefunction-based techniques like MP2/RI-MP2 and CCSD(T)/DF-CCSD(T) have seen successful optimization for GPU processing. Our work in GPU4PySCF includes preliminary implementations for MP2 and CCSD methods, demonstrating the feasibility and benefits of using GPUs for these computationally intensive tasks. References such as~\cite{barca_rimp2,bykov_rimp2,Kwack_rimp2} for MP2/RI-MP2 and~\cite{datta_dfccsd,kim_ccsd} for CCSD(T)/DF-CCSD(T) highlight the ongoing efforts and achievements in this area.

{\bf Direct SCF \& Other Integral Schemes}. The Direct Self-Consistent Field (SCF) method has been particularly explored for GPU acceleration due to its minimal memory requirements and scalability with system size, especially when screening techniques are applied. The efficiency of Direct SCF is largely dependent on the integral computation strategy. As described in Ref.~\cite{gpu4pyscf-caltech} the start of the GPU4PySCF project was built around Direct SCF using a Rys quadrature scheme to compute the 4-center two-electron integrals. The Rys quadrature scheme was selected mainly because of its simplicity for implementing various operators and derivatives. In future work, it may be worth revisiting the integral scheme for direct SCF calculations.

{\bf Periodic Boundary Conditions (PBC)}. The PBC module in PySCF extends its capabilities to simulate extended systems using Gaussian basis sets. The main challenges in PBC calculations can be attributed to the infinite number of repeated images caused by periodicity. This leads to high computational costs and numerical instability. Although sophisticated integral screening and evaluation schemes have been implemented in the CPU code, their techniques may not be directly transferable to GPU-based programs. New techniques are under development to refine integral screening and evaluation for GPU-optimized PBC simulations, as mentioned in~\cite{sun2023various}.

{\bf Multi-GPUs}. The advent of data-center GPUs equipped with multi-GPU configurations significantly enhances data transfer speeds via technologies like NVIDIA's NVLink or AMD's xGMI, offering a substantial improvement over traditional CPU-GPU data transfer rates. This technique not only accelerates the computation of quantum chemistry problems but also enables the efficient handling of tasks that require large amounts of GPU memory. Multi-GPU implementations perhaps can achieve greater parallel efficiency than 100\% for large molecules.

{\bf Mixed Precision}. The mixed-precision computing approach capitalizes on the differential computational capabilities of GPUs, balancing single and double precision to optimize performance without compromising accuracy. In the context of Direct SCF algorithms, error estimation via Schwarz inequality allows for the strategic use of precision levels. This technique reduces computational overhead by performing most calculations in single precision in later SCF iterations. Such strategies, as investigated in~\cite{mixed_rys, terachem2011}, showcase the ongoing innovation in leveraging GPU capabilities to improve quantum chemistry computations, focusing on maximizing performance and computational efficiency.

{\bf Low-level Integration with other PySCF-based Packages}. Due to fundamental differences between the underlying CPU algorithms and their corresponding GPU algorithms, the low-level APIs of PySCF and GPU4PySCF are not fully compatible. As a result, packages relying on low-level PySCF APIs cannot be directly accelerated with GPU4PySCF. Additionally, we are aware of other groups working on PySCF-based GPU implementations. Other designs could use JAX or Torch as the linear algebra backend. Defining the interfaces between these packages remains an area of active exploration.

\section*{Acknowledgements}
The authors would like to thank the organizers of the 1st annual PySCF Developers Meeting. The initial results for GPU acceleration of PySCF were presented at that meeting by Dr. Qiming Sun before the public release of the GPU4PySCF repository. The authors thank all the attendees for their discussions. After the meeting, Dr. Garnet Chan and Dr. Qiming Sun initiated the public GPU4PySCF repository onGitHub, and the Caltech team developed the GPU accelerated 4-center GTO integrals and direct SCF scheme that formed the initial version of GPU4PySCF. This initial version is described further in Ref.~\cite{gpu4pyscf-caltech}. The authors also would like to thank Dr. John Herbert from The Ohio State University for providing the implementation details of CHELPG charge in Q-Chem 6.1.

\section*{Author Contributions}
%From 2023-2024, {\bf Qiming Sun} worked as a Third-Party Associate for ByteDance Inc. 
{\bf Xiaojie Wu} and {\bf Qiming Sun} designed the framework and APIs in this work; built CI; analyzed the performance; conceptualized the paper and planned the application projects. {\bf Qiming Sun} contributed this work when he worked as an independent researcher. {\bf Xiaojie Wu} optimized the framework of DFT; implemented density fitting, gradient, Hessian, and solvent models; performed benchmarks of DF modules with calculating solvation free energy, transition state search. {\bf Zhichen Pu} developed open-shell HF/DFT, CHELPG charge, and NMR. {\bf Tianze Zheng} and {\bf Wen Yan} performed the applications in torsion scan and dimer interaction energies. {\bf Wenzhi Ma} performed the applications of Fukui functions and ESP. {\bf Zhichen Pu}, {\bf Sheng Gong}, {\bf Yumin Zhang} and {\bf Weihao Gao} studied Lithium ion solvation structure and electrochemical stability windows of electrolytes. {\bf Weiluo Ren} and {\bf Xiang Li} incorporated GPU4PySCF with neural networks. {\bf Yu Xia}, {\bf Zhengxiao Wu}, and {\bf Mian Huo} contributed the engineering improvements. All authors contributed the paper draft. 

\appendix
\section{Dataset for Benchmarks}
The information of the dataset used in Section \ref{sec:benchmark} is as follow:
\label{sec:dataset}
\begin{table}[!htp]
    \centering
    \begin{tabular}{lll}
\hline
Molecule name & Number of atoms & Elements\\
\hline
Vitamin C & 20 & H,C,O\\
Inosine & 31 & H,C,O,N\\
Bisphenol A & 33 & H,C,O\\
Mg Porphin & 37 & H,C,N,Mg\\
Penicillin V & 42 & H,C,O,N,S\\
Ochratoxin A & 45 & H,C,O,N,Cl\\
Cetirizine & 52 & H,C,O,N,Cl\\
Tamoxifen & 57 & H,C,O,N\\
Raffinose & 66 & H,C,O\\
Sphingomyelin & 84 & H,C,O,N,P\\
Azadirachtin & 95 & H,C,O\\
Taxol & 113 & H,C,O\\
Valinomycin & 168 & H,C,O,N\\
\hline
\end{tabular}
\caption{The information of constructed dataset.}
\label{tab:dataset}
\end{table}

\newpage
\section{Accuracy for Different XC Functionals}
\label{sec:accuracy_xc}
The following calculations are using def2-TZVPP, def2-universal-JKFIT, and (99,590) grids. The energy, gradient, and Hessian difference are measured in 2-norm. The units are Hartree, Hartree/Bohr, and Hartree/Bohr$^2$ respectively.
\begin{table}[!htp]
    \centering
    \begin{tabular}{c|ccc}
\hline
Molecule names	& Energy & Gradient & Hessian\\
\hline
Vitamin C	& 1.30E-06 &	3.35E-05&	1.26E-03\\
Inosine	& 5.55E-08 &	3.44E-05&	1.19E-03\\
Bisphenol A	& 4.99E-06 &	6.39E-05&	1.51E-03\\
Mg Porphin	& 9.23E-06 &	4.10E-05&	1.34E-03\\
Penicillin V &	2.48E-06 &	1.12E-04&	2.18E-03\\
Ochratoxin A &	9.07E-06 &	9.93E-05&	2.33E-03\\
Cetirizine	& 1.12E-05 &	1.30E-04&	4.77E-03\\
Tamoxifen	& 7.96E-06 &	8.98E-05&	1.77E-03\\
Raffinose	& 8.97E-06 &	2.15E-04&	2.96E-03\\
Sphingomyelin &	1.69E-05&	1.27E-04&	4.10E-03\\
\hline
    \end{tabular}
    \caption{Difference between GPU4PySCF and Q-Chem 6.1, using PBE XC functional.}
    \label{tab:PBE}
\end{table}
%
%\begin{table}[!htp]
%    \centering
%    \begin{tabular}{c|ccc}
%\hline
%Molecule names	& Energy & Gradient & Hessian\\
%\hline
%Vitamin C &	7.31E-06 &	1.36E-04 &	6.59E-03\\
%Inosine	& 2.65E-05 &	2.38E-04 &	1.04E-02\\
%Bisphenol A &	1.13E-05 &	2.17E-04 &	7.08E-03\\
%Mg Porphin &	2.76E-04 &	5.48E-04 &	3.09E-02\\
%Penicillin V &	6.45E-06 &	5.04E-04 &	1.16E-02\\
%Ochratoxin A &	1.70E-05 &	4.98E-04 &	1.60E-02\\
%Cetirizine &	2.70E-05 &	3.64E-04 &	1.10E-02\\
%Tamoxifen &	2.17E-05 &	3.65E-04 &	1.37E-02\\
%Raffinose &	2.96E-05 &	8.44E-04 &	1.81E-02\\
%Sphingomyelin &	3.35E-06 &	3.13E-04 &	1.94E-02\\
%\hline
%    \end{tabular}
%    \caption{Difference between GPU4PySCF and Q-Chem 6.1, using M06 XC functional.}
%    \label{tab:M06}
%\end{table}

\begin{table}[!htp]
    \centering
    \begin{tabular}{c|ccc}
\hline
Molecule names	& Energy & Gradient & Hessian\\
\hline
Vitamin C	& 5.55E-07 &	2.74E-05 &	1.04E-03\\
Inosine	& 8.17E-07 &	2.76E-05 &	9.62E-04\\
Bisphenol A &	4.61E-06 &	5.21E-05 &	1.17E-03\\
Mg Porphin	& 8.28E-06 &	3.82E-05 &	1.20E-03\\
Penicillin V &	1.86E-06 &	8.76E-05 &	1.69E-03\\
Ochratoxin A &	7.67E-06 &	6.80E-05 &	1.77E-03\\
Cetirizine &	8.40E-06 &	9.09E-05 &	3.56E-03\\
Tamoxifen &	4.77E-06 &	7.09E-05 &	1.29E-03\\
Raffinose &	6.43E-06 &	1.62E-04 &	2.22E-03\\
Sphingomyelin &	1.52E-05 &	9.50E-05 &	3.28E-03\\
\hline
    \end{tabular}
    \caption{Difference between GPU4PySCF and Q-Chem 6.1, using B3LYP XC functional.}
    \label{tab:B3LYP}
\end{table}

\begin{table}[!htp]
    \centering
    \begin{tabular}{c|cc}
\hline
Molecule names	& Energy & Gradient\\
\hline
Vitamin C &	1.91E-06 &	2.45E-05\\
Inosine	& 3.32E-06 &	3.96E-05\\
Bisphenol A &	3.07E-06 &	6.54E-05\\
Mg Porphin	& 5.40E-06 &	4.38E-05\\
Penicillin V &	3.35E-06 &	9.26E-05\\
Ochratoxin A &	5.29E-06 &	7.22E-05\\
Cetirizine &	1.54E-05 &	7.74E-05\\
Tamoxifen &	3.78E-06 &	7.54E-05\\
Raffinose &	7.88E-07 &	1.28E-04\\
Sphingomyelin &	1.76E-05 &	9.69E-05\\
\hline
    \end{tabular}
    \caption{Difference between GPU4PySCF and Q-Chem 6.1, using $\omega$B97M-V XC functional, Hessian is not supported yet.}
    \label{tab:wb97m-v}
\end{table}

\newpage
\section{Accuracy for Different Basis Sets}
\label{sec:accuracy_basis}
The following calculations are using B3LYP, def2-universal-JKFIT, (99,590) grids. The energy, gradient, and Hessian difference are measured in 2-norm. The units for energy, gradient, and Hessian are Hartree, Hartree/Bohr and Hartree/Bohr$^2$ respectively. 
\begin{table}[!htp]
    \centering
    \begin{tabular}{c|ccc}
\hline
Molecule names	& Energy & Gradient & Hessian\\
\hline
Vitamin C	& 1.14E-07	&2.65E-05	&9.88E-04\\
Inosine	& 1.06E-05	&2.71E-05	&1.03E-03\\
Bisphenol A	& 1.28E-06	&4.68E-05	&1.06E-03\\
Mg Porphin &	6.26E-06	&3.97E-05	&1.01E-03\\
Penicillin V&	5.18E-05	&8.46E-05	&1.65E-03\\
Ochratoxin A&	4.72E-05	&5.39E-05	&1.61E-03\\
Cetirizine&	2.46E-05	&9.08E-05	&3.18E-03\\
Tamoxifen&	3.86E-05	&6.43E-05	&1.28E-03\\
Raffinose&	7.80E-05	&1.41E-04	&2.18E-03\\
Sphingomyelin&	8.46E-05	&8.99E-05	&3.14E-03\\
\hline
    \end{tabular}
    \caption{Difference between GPU4PySCF and Q-Chem 6.1, using 6-31G basis.}
    \label{tab:6-31G}
\end{table}

\begin{table}[!htp]
    \centering
    \begin{tabular}{c|ccc}
\hline
Molecule names	& Energy & Gradient & Hessian\\
\hline
Vitamin C	&6.31E-07&	2.52E-05&	9.24E-04\\
Inosine	& 1.25E-05&	2.61E-05&	9.82E-04\\
Bisphenol A	& 9.31E-07&	4.68E-05&	9.89E-04\\
Mg Porphin	& 7.19E-06&	3.15E-05&	9.59E-04\\
Penicillin V &	3.71E-05&	8.31E-05&	1.52E-03\\
Ochratoxin A &	3.06E-05&	5.06E-05&	1.50E-03\\
Cetirizine	& 2.26E-05&	8.67E-05&	2.93E-03\\
Tamoxifen	& 1.73E-05&	6.38E-05&	1.27E-03\\
Raffinose	& 5.60E-05&	1.32E-04&	2.13E-03\\
Sphingomyelin&	4.12E-05&	8.31E-05&	3.13E-03\\
\hline
    \end{tabular}
    \caption{Difference between GPU4PySCF and Q-Chem 6.1, using def2-SVP basis.}
    \label{tab:def2-SVP}
\end{table}

\begin{table}[!htp]
    \centering
    \begin{tabular}{c|ccc}
\hline
Molecule names	& Energy & Gradient & Hessian\\
\hline
Vitamin C	& 5.55E-07 &	2.74E-05 &	1.04E-03\\
Inosine	& 8.17E-07 &	2.76E-05 &	9.62E-04\\
Bisphenol A	& 4.61E-06 & 	5.21E-05 &	1.17E-03\\
Mg Porphin	& 8.28E-06 &	3.82E-05 &	1.20E-03\\
Penicillin V &	1.86E-06 &	8.76E-05 &	1.69E-03\\
Ochratoxin A &	7.67E-06 &	6.80E-05 &	1.77E-03\\
Cetirizine	& 8.40E-06 &	9.09E-05 &	3.56E-03\\
Tamoxifen	& 4.77E-06 &	7.09E-05 &	1.29E-03\\
Raffinose	& 6.43E-06 &	1.62E-04 &	2.22E-03\\
Sphingomyelin &	1.52E-05&	9.50E-05 &	3.28E-03\\
\hline
    \end{tabular}
    \caption{Difference between GPU4PySCF and Q-Chem 6.1, using def2-TZVPP basis set.}
    \label{tab:def2-TZVPP}
\end{table}

\begin{table}[!htp]
    \centering
    \begin{tabular}{c|ccc}
\hline
Molecule names	& Energy & Gradient & Hessian\\
\hline
Vitamin C	& 5.23E-07 &	2.79E-05 &	1.04E-03\\
Inosine	& 9.11E-07 &	2.75E-05 &	9.63E-04\\
Bisphenol A	& 4.78E-06 &	5.23E-05 &	1.17E-03\\
Mg Porphin	& 8.28E-06 &	3.83E-05 &	1.21E-03\\
Penicillin V &	4.23E-06 &	8.74E-05 &	1.70E-03\\
Ochratoxin A &	3.08E-06 &	6.81E-05 &	1.77E-03\\
Cetirizine	& 6.85E-06 &	9.10E-05 &	3.56E-03\\
Tamoxifen	& 3.36E-06 &	7.12E-05 &	1.29E-03\\
Raffinose	& 2.57E-06 &	1.61E-04 &	2.22E-03\\
Sphingomyelin &	6.98E-06&	9.57E-05 &	3.29E-03\\
\hline
    \end{tabular}
    \caption{Difference between GPU4PySCF and Q-Chem 6.1, using def2-TZVPD basis set.}
    \label{tab:def2-TZVPD}
\end{table}

\newpage
\section{Accuracy for Solvent Models}
\label{sec:accuracy_solvent}
All of the following calculations are using B3LYP, def2-TZVPP, def2-universal-JKFIT, (99,590) grids. The energy, gradient, and Hessian difference are measured in 2-norm. The corresponding units are Hartree, Hartree/Bohr and Hartree/Bohr$^2$. Since Q-Chem 6.1 do not support PCM models for the density fitting scheme, we use the regular SCF scheme as the reference. The following tables reflects the density fitting errors.
\begin{table}[!htp]
    \centering
    \begin{tabular}{c|ccc}
\hline
Molecule names	& Energy & Gradient & Hessian\\
\hline
Vitamin C &	1.38E-04 &	1.20E-04 &	1.14E-03\\
Inosine	& 2.08E-04	 & 1.39E-04	 & 1.18E-03\\
Bisphenol A &	1.99E-04 &	1.78E-04 &	1.36E-03\\
Mg Porphin &	1.40E-04 &	1.35E-04 &	1.33E-03\\
\hline
    \end{tabular}
    \caption{Difference between GPU4PySCF and Q-Chem 6.1, using C-PCM solvent model.}
    \label{tab:C-PCM}
\end{table}

\begin{table}[!htp]
    \centering
    \begin{tabular}{c|ccc}
\hline
Molecule names	& Energy & Gradient & Hessian\\
\hline
Vitamin C &	1.38E-04 &	1.20E-04 &	1.14E-03\\
Inosine	& 2.08E-04 &	1.39E-04 &	1.18E-03\\
Bisphenol A &	1.99E-04 &	1.78E-04 &	1.36E-03\\
Mg Porphin &	1.40E-04 &	1.35E-04 &	1.33E-03\\
\hline
    \end{tabular}
    \caption{Difference between GPU4PySCF and Q-Chem 6.1, using IEF-PCM solvent model}
    \label{tab:IEF-PCM}
\end{table}
\bibliographystyle{unsrt}
\bibliography{main}

\end{document}